 \documentclass[aps,prb,twocolumn,superscriptaddress,showpacs,amsmath,amssymb]{revtex4}
\usepackage{graphicx}% Include figure files
\usepackage{dcolumn} % Align table columns on decimal point
\usepackage{bm}      % bold math
\usepackage{verbatim}%

\begin{document}

%\preprint{}
%\preprint{APS/123-QED}

%------------------------- Title --------------------------%
\title
{Solid-Liquid  Phase Diagrams for Binary Metallic Alloys:  Adjustable Interatomic Potentials}
%Solid-Liquid Binary Phase Diagrams for generic embedded-atom model metal 

%----------------- Author and Affiliation -----------------%

% The \affiliation command should follow the other information
%\author{}
%\thanks{}
%\affiliation{}

\author{ H.-S. Nam }
\email
{hnam@princeton.edu}

%\altaffiliation
%{Present address:xxx}
\affiliation
{
Department of Mechanical and Aerospace Engineering, Princeton University, Princeton, New Jersey 08544
}

\author{ M. I. Mendelev }

\affiliation
{
Materials and Engineering Physics, Ames Laboratory, Ames, Iowa 50011
}

\author{ D. J. Srolovitz }

\affiliation
{
Department of Mechanical and Aerospace Engineering, Princeton University, Princeton, New Jersey 08544
}
\affiliation
{
Department of Physics, Yeshiva University, New York, NY 10033
}


%-------------------------- Date --------------------------%
\date{\today}% It is always \today, today,

%------------------------ Abstract ------------------------%
\begin{abstract}

We develop an approach to determining LJ-EAM potentials for alloys and use these to determine the solid-liquid phase diagrams for binary metallic alloys using Kofke's Gibbs-Duhem integration technique combined with semigrand canonical Monte Carlo simulations. We demonstrate that it is possible to produce a wide-range of experimentally observed binary phase diagrams (with no intermetallic phases) by reference to the atomic sizes and cohesive energies of the two elemental materials. In some cases, it is useful to employ a single adjustable parameter to adjust the phase diagram (we provided a good choice for this free parameter).  Next, we perform a systematic investigation of the effect of relative atomic sizes and cohesive energies of the elements on the binary phase diagrams.  We then show that this approach leads to good agreement with several experimental binary phase diagrams.  The main benefit of this approach is not the accurate reproduction of experimental phase diagrams, but rather to provide a method by which material properties can be continuously changed in simulation studies.  This is one of the keys to the use of atomistic simulations to understand mechanisms and properties in a manner not available to experiment.

\end{abstract}

%---------------------- PACS number -----------------------%
% 81. Materials science
% 81.30.Bx Phase diagrams of metals and alloys
%
% 82. Physical chemistry and chemical physics
% 82.60.Lf Thermodynamics of solutions
%
% 02. Mathematical methods in physics
% 02.70.Uu Applications of Monte Carlo methods
%
% 64. Equations of state, phase equilibria, and phase transitions
% 64.70.Dv Solid-liquid transitions
%


%----------------------- Key Words ------------------------%
%\keywords{}

\maketitle

%%%%%%%%%%%%%%%%%%%%%%%%%%%%%%% Body of Paper %%%%%%%%%%%%%%%%%%%%%%%%%%%%%%%%%%

\section{\label{sec:level1} Introduction}

Atomic-scale simulations have become an indispensable tool for modern materials research.  For example, molecular dynamics (MD) and Monte Carlo (MC) simulations are routinely used to investigate phenomena that are not easily accessible via experiment or to interpret experimental results.\cite{Ohno:CMS, Frenkel:UnderstandingMS}  The fundamental input to such simulations is a description of the interactions between atoms.  While first principles methods accurately describe atomic bonding through quantum mechanical treatments, they are usually limited to a relatively small number of atoms.  Semi-empirical or empirical interatomic descriptions are often motivated by quantum mechanical ideas but represent different materials through parameterization schemes in which the constants are fit to experimental (and/or first principles) data.  While this approach has significant problems when the resultant potentials are employed under conditions for which they were not fitted, they can provide accurate results when applied carefully.  Such potentials have the advantage that they can be easily used for simulations involving a very large number of atoms (currently up to $10^9$).  Since few interesting materials are pure, we focus on potentials for metallic alloys in this paper.  In order to determine the utility of potentials for alloys, we should insure that the potentials lead to the correct phases at temperatures and compositions of interest.  In this paper, we describe the determination of binary phase diagrams for a particularly flexible choice of potentials for metallic alloys.

While there has been a long tradition of modeling materials using potentials that can be adjusted to represent different types of materials, such readily adjustable potentials are commonly pairwise.  As such, they do not provide a reasonable description of metals (e.g., surface relaxation). Ideally, we will be able to easily adjust a potential to give the desired metallic phase diagram.  

Examples of easily adjustable pairwise interatomic potentials include the Lennard-Jones\cite{LennardJones:LJpotential} (LJ) and Morse\cite{Morse:Morsepotential} potentials. Because of their simplicity and applicability to systems with a wide range of properties, these potentials have been widely used, both for elemental and multi-component systems.\cite{Jensen:LJexample, Broughton:LJexample}  However, such potentials are only realistic for very simple materials, such as noble gases. In other materials, bonding is more complex.  For example, for metals and alloys, it is well known that many-body effects play an important role.\cite{Vitek:BeyondPair, Carlsson:BeyondPair}  For such materials, potentials of the embedded atom method\cite{Daw:EAM} (EAM) type are widely used. Holian \emph {et al.}\cite{Holian:LJEAM} proposed an extension of the Lennard-Jones potential that allows for many body interactions, like in EAM type potentials.  This idea was further pursued by Baskes\cite{Baskes:LJEAMPRL, Baskes:LJEAMPRSL, Baskes:PhaDiaEAM} to treat a broad range of metallic systems including alloys resulting in a potential known as the Lennard-Jones, Embedded Atom Method (LJ-EAM) potential.  This potential represents a readily adjustable description of atomic interactions in metals and metallic alloys.  Because of the relatively simple analytical form of the adjustable LJ-EAM potential, it can be used as a description of atomic interactions in systems with different, controllable thermodynamics properties.  As such, it provides a ready means to test the influence of thermodynamic properties on materials behavior without the complexity of first principles approaches or all of the oversimplifications inherent in pairwise potentials. 

In this paper, we focus on the systematic determination of binary phase diagrams for LJ-EAM potentials.  The ability to predict phase diagrams for such adjustable interatomic potentials is an important first step in identifying a set of atomic interactions required to reproduce the requisite phases needed to describe specific phenomena and systems at non-zero temperature.  This is also important for determining which thermodynamic properties are important in particular type of phenomena observed experimentally.  We were motivated to pursue this study through our own attempts to perform molecular dynamics simulations of liquid metal embrittlement in alloys.\cite{Joseph:LME}  There have been several contradictory suggestions as to what type of thermodynamic behavior is necessary for this phenomenon to occur.  Therefore, we specifically consider the solid-liquid regions of the phase diagrams for a wide range of metallic binary alloys.  We focus on two main parameters in describing the alloys - relative atomic size and the strength of the chemical bonds.  Of course, this is not the first attempt to systematically describe binary phase diagrams from an atomistic view.  Earlier attempts have examined the phase diagrams of hard sphere\cite{Kranendonk:HardSphere} and Lennard-Jones materials.\cite{Vlot:PhaDiaLJ, Hitchcock:JCP}  The methods employed to determine the phase behavior range from density functional theory\cite{Octoby:DFT} to Gibbs-Duhem integration methods.\cite{Kofke:GibbsDuhemMolPhys, Kofke:GibbsDuhemJCP}  We combine LJ-EAM potentials, molecular dynamics, and Gibbs-Duhem integration methods in a Monte Carlo framework to develop a systematic understanding of the relationship between potential properties and the solid/liquid phase diagram.

\section{\label{sec:level2} Potentials}

In this section, we outline the form of the LJ-EAM model used in this work.  For a binary material described by classical LJ pair potentials, the interatomic potential between atoms $i$ and $j$ takes the form:
\begin{equation}
\label{eq:LJpair}
\phi_{s_i s_j}^{LJ} (r) = 4 \epsilon_{s_i s_j} \left [ \left ( \frac {\sigma_{s_i s_j}} {r} \right )^{12} - \left ( \frac {\sigma_{s_i s_j}} {r} \right )^{6} \right ] ,
\end{equation}
where $\epsilon_{s_i s_j}$ and $\sigma_{s_i s_j}$ are the attractive well depth and the diameter for the LJ potential describing the interactions between species $ s_i $ and $ s_j $ ($s=A$ or $B$).
The total energy of a binary LJ-EAM system is given by the usual EAM form:
\begin{equation}
\label{eq:EAM}
E ~ = ~ \sum_{i} \left [ F_{s_i} ( \bar \rho_i )
~ + ~ \frac {1} {2} \sum_{j \neq i} \phi_{s_i s_j}  ( r_{ij} ) \right ] ,
\end{equation}
where $ F_{s_i} ( \bar \rho_i )$ is the embedding energy and $\phi_{s_i s_j} ( r_{ij} )$ is the pair interaction between atoms $i$ and $j$ separated by a distance $r_{ij}$.
The embedded function is commonly chosen as:\cite{Baskes:LJEAMPRL}
\begin{subequations}
\begin{equation}
\label{eq:EAMf}
F_{s_i} ( \bar \rho_{i} ) ~ = ~ \frac {1} {2} \widehat{A}_{s_i} Z_{1} \epsilon_{s_i s_i} \bar \rho_{i} \left [ \ln \left ( \bar \rho_{i} \right ) - 1 \right ] ,
\end{equation}
where the electron density at the site of atom is 
\begin{equation}
\bar \rho_i ~ = ~  \frac {1}{Z_1} \sum_{j \neq i} \rho_{j} ( r_{ij} )
\end{equation}
and
\begin{equation}
\label{eq:EAMrho}
\rho_{j} ( r_{ij}  ) ~ = ~ \exp [ \hat{\beta}_{s_j} ( \frac {r_{ij}} {\sqrt[6]{2} \sigma_{s_i s_j}} - 1 ) ] .
\end{equation}
\end{subequations}
Here, the dimensionless parameter $\widehat{A}_{s}$ represents the strength of the many-body term, the parameter $\hat{\beta}_{s}$ quantifies the distance over which the electron density decays away from an atom position, and $Z_1$ is the coordination number of the reference state (e.g., face centered cubic). 

We can combine the EAM form of the total energy appropriate for metals with the convenience of the adjustable LJ pair potential by choosing the pairwise term in Eq.~(\ref{eq:EAM}) in such a way that the total energy of the reference structure as a function of dilation is described by a LJ potential.  If we include interactions up to second-nearest-neighbors, the like atom pair potential $\phi_{AA} (r)$ for species $A$ is given by
\begin{subequations}
\begin{equation}
\label{eq:EAMphiHomo}
\phi_{AA} (r) + \left ( \frac {Z_2} {Z_1} \right ) \phi_{AA} (ar) ~=~ \psi_{A} (r)
\end{equation}
or
\begin{eqnarray}
\phi_{AA} (r) ~=~ \psi_{A} (r) + \sum_{n=1}^{N} { \left [ (-1)^{n} \left ( \frac {Z_2} {Z_1} \right )^{n} \psi_{A} (a^{n} r) \right ] } ,
\end{eqnarray}
where 
\begin{eqnarray}
\psi_{A} (r) ~=~ \phi_{AA}^{LJ} (r) - \left ( \frac {2} {Z_1} \right ) F_{A} (\bar \rho_{A}^{0} (r) ) 
\end{eqnarray}
and
\begin{eqnarray}
\bar \rho_{A}^{0} (r) ~=~ \rho_{A} (r) + \left ( \frac {Z_2} {Z_1} \right ) \rho_{A} (ar) .
\end{eqnarray}
\end{subequations}
Here, $Z_2$ is the number of second-nearest-neighbor atoms and $a$ is the ratio of the second- to first-nearest-neighbor distance. The summation of $N$ terms is carried out until $\phi_{AA}(r)$ converges.

Because this potential was fitted to a LJ-form, it has only four adjustable parameters, $\widehat{A}_{A}$, $\hat{\beta}_{A}$, $\epsilon_{A}$ ($\equiv\epsilon_{AA}$), and $\sigma_{A}$ ($\equiv\sigma_{AA}$) for the single component material, $A$.  This potential form can be easily extended to multi-component systems. For EAM binary alloys, we must fix seven functions $\rho_{A}(r)$, $\rho_{B}(r)$, $F_{A}(r)$, $F_{B}(r)$, $\phi_{AA}(r)$, $\phi_{BB}(r)$ and $\phi_{AB}(r)$.  The first six of these are transferable from the two monatomic systems. Following Baskes and Stan,\cite{Baskes:PhaDiaEAM} we can obtain the remaining function, $\phi_{AB} (r)$, by fitting to a particular alloy structure.  Like them, we focus on the ordered $L1_0$ compound (after correcting a small error in Ref.~\onlinecite{Baskes:PhaDiaEAM}), as described in Appendix A:
\begin{eqnarray}
\label{eq:EAMphiHetro}
\phi_{AB} (r) =&&\phi_{AB}^{LJ} (r)
- \frac {1} {8} \left [ F_{A} (\bar \rho_{A}^{L1_0} (r) ) + F_{B} (\bar \rho_{B}^{L1_0} (r) ) \right ]
\nonumber\\
&&- \frac {1} {4} \left [ \phi_{AA} (r) + \phi_{BB} (r) - \phi_{AA}^{LJ} (r) - \phi_{BB}^{LJ} (r) \right ]
\nonumber\\
&&- \frac {3} {8} \left [ \phi_{AA} (ar) + \phi_{BB} (ar) \right ] .
\end{eqnarray}
If we set $\widehat{A}_{A}=0$ and $\widehat{A}_{B}=0$,  $\phi_{AA}$ and $\phi_{AB}$ take exactly the same form as the LJ potential.

$\phi_{AB}(r)$ is fully determined by Eq.~(\ref{eq:EAMphiHetro}), except for $\phi_{AB}^{LJ}(r)$.  Therefore, we need two parameters, $\epsilon_{AB}$ and $\sigma_{AB}$, to describe the cross interaction between species $A$ and $B$ (in addition to the 8 parameters needed to describe pure $A$ and pure $B$). 
The cross-species interaction parameters $\epsilon$ and $\sigma$ are, therefore, the knobs we use to control alloy properties. The Lorentz-Bertholet mixing rules\cite{Rowlinson:LBrule} are widely used for obtaining alloy parameters in LJ systems; i.e.,
$ \sigma_{AB} = \left ( \sigma_{A} + \sigma_{B} \right )/2 $ and $ \epsilon_{AB} = \sqrt { \epsilon_{A} \epsilon_{B} } $.
However, applying the Lorentz-Berthelot mixing rules to determine $\epsilon_{AB}$ and $\sigma_{AB}$ results in phase diagrams that do not correspond to those observed for metallic alloys (i.e., the solid/liquid two phase region is quite square with very limited solubility in the solid and the liquid). Even though the Lorentz-Berthelot mixing rules may be appropriate for LJ potentials, there is no reason to expect this to be true for LJ-EAM potentials.  This is because in the LJ-EAM potentials the pairwise interaction represents just part of the bonding and it is the total LJ-EAM potential that must reproduce the LJ potential for dilation.  Therefore, to construct the pairwise interaction term in LJ-EAM, 
we apply the Lorentz-Berthelot mixing rule to $\phi_{AB}(r)$ in Eq.~(\ref{eq:EAMphiHetro}),  rather than to $\phi^{LJ}_{AB}(r)$ itself.  Doing this, we find the well depth $\epsilon_{AB}$ should be
\begin{eqnarray}
\label{eq:NewRule}
\epsilon_{AB} = 
&&\frac {1} {2} \left ( \widehat{A}_{A} \epsilon_{A} + \widehat{A}_{B} \epsilon_{B} \right ) 
\nonumber\\
&&+ \sqrt { |(1 - \widehat{A}_{A}) \epsilon_{A} (1 - \widehat{A}_{B}) \epsilon_{B}| } .
\end{eqnarray}
The derivation of this rule is given in Appendix B.

Although LJ-EAM potentials are not as widely used as the classic LJ potential, the LJ-EAM potential has already been widely applied in the literature.\cite{Baskes:LJEAMPRL, Baskes:LJEAMPRSL, Baskes:PhaDiaEAM, Yurtsever:LJEAMcluster, Wang:LJEAMmelting, Baskes:LJEAMbinary, Srolovitz:StressMorphology} The equilibrium structure of a LJ crystal at 0 K is face centered cubic (FCC) and the melting point depends solely on the cohesive energy and the interaction range of the potential.\cite{Pryde:LiquidState} However, both the ground state structure and melting temperature of single component LJ-EAM system are strongly dependent on the two many-body interaction parameters $\widehat{A}$ and $\hat{\beta}$.\cite{Baskes:LJEAMPRL}  As the strength of many-body interaction $\widehat{A}$ increases, the melting points of the pure elements decrease although the cohesive energy remains unchanged.\cite{Baskes:LJEAMPRL}  Experimentally, most FCC metals exhibit normalized melting temperatures $k_{B}T_{m}/E_{0}$ in the range from 0.025 to 0.04, where $k_{B}$ is the Boltzmann constant and $E_{0}$ the cohesive energy of the solid at zero pressure. If we set $\widehat{A}=0.7 $ and $\hat{\beta}=7 $ in the LJ-EAM potentials for both components, pure $A$ and pure $B$ will be FCC at $T=0$ and the normalized melting points fall within this temperature range for FCC metals.\cite{Baskes:LJEAMPRSL} 

The melting point of an elemental FCC solid is strictly proportional to the well depth $\epsilon$ (for any choice of $\widehat{A}$ and $\hat{\beta}$). For the reference element $A$, we set $\epsilon = 0.6$ eV and $\sigma = 2.5$ {\AA} with which the melting point $T_{m}=1405$ K was obtained. However, this choice of $\epsilon$ and $\sigma$ is rather arbitrary, since only the ratios of well-depths ${\epsilon_{B}} / {\epsilon _{A}}$ and diameters ${\sigma_{B}} / {\sigma_{A}}$ are relevant in determining the binary phase diagrams for LJ-EAM potentials. The potential interactions were truncated between the second- and third-nearest-neighbors such that the relaxed cohesive energy is $E_0=6.0 \epsilon$ for the FCC reference state [see Eq.~(\ref{eq:EAMphiHomo})]. 

Table~\ref{tab:table1} shows a comparison of the basic properties of the LJ-EAM material with those of several FCC noble metals. For the LJ-EAM parameters employed in Table~\ref{tab:table1}, the LJ-EAM potential yields reasonable properties for the FCC metals except for the bulk modulus. Unlike other properties, the bulk modulus is solely determined by the second derivative of the energy as a function of lattice dilation (regardless of the many-body interactions). Since the LJ-EAM potential was fitted to a LJ-form (by definition), both the LJ and LJ-EAM models yield similar bulk moduli.  The bulk modulus of LJ systems is known to be too large compared with the FCC metals.\cite{Baskes:LJEAMPRL} One could obtain better bulk modulus values by fitting the pair interactions to another form, e.g., the universal binding energy relation.\cite{Rose:EAMUniversalFeatures} 

\begin{table*}[!tbp]
\caption
{\label{tab:table1}
A comparison of several key properties for the LJ-EAM potential model (for $\widehat{A}=0.7$, $\hat{\beta}=7$) and several FCC metals.
All properties are normalized by using $E_{0}(=6\epsilon)$, $r_{0}(=\sqrt[6]{2}\sigma)$, and $\Omega(=r^{3}_{0}/\sqrt{2})$.
}
\begin{ruledtabular}
\begin{tabular}{lccccccc}
property & normalized quantity & LJ & LJ-EAM & Cu & Ag & Au & FCC metals \\
\hline
bulk modulus            & $B \Omega / E_{0}$              & 8.5   & 8.5   & 3.0   & 3.92  & 5.05  & 2.0-6.0     \\
Melting point           & $k_{B} T_{\rm{m}}/E_{0}$        & 0.082 & 0.033 & 0.033 & 0.037 & 0.029 & 0.024-0.041 \\
Cauchy discrepancy      & $(c_{12}-c_{44})/B$             & 0     & 0.48  & 0.30  & 0.425 & 0.69  &-0.4-0.7     \\
vacancy formation energy &  $E^{f}_{v} / E_{0}$           & 1.0   & 0.31  & 0.37  & 0.39  & 0.23  & 0.2-0.5     \\
(100) surface energy    & $E_{(100)} r^{2}_{0} / E_{0}$   & 0.67  & 0.26  & 0.24  & 0.23  & 0.22  & 0.15-0.25   \\
\end{tabular}
\end{ruledtabular}
\end{table*}

\section{\label{sec:level3} Gibbs-Duhem integration method }

There are several computational methods that can be used to determine the phase diagram of a system described by any choice for the interatomic potentials.\cite{Panagiotopoulos:GibbsEnsemble, Kofke:GibbsDuhemMolPhys}  Here, we directly construct the phase diagram using the Gibbs-Duhem integration technique proposed by Kofke.\cite{Kofke:GibbsDuhemMolPhys, Kofke:GibbsDuhemJCP}  In this method, two or more coexisting phases are simulated (semi-grand canonical Monte Carlo simulations) independently at the same temperature and pressure. Once a single point of the coexistence curve between two phases is known, the rest of the curve can be computed (without any free energy calculations) by integrating the equivalent of the Clausius-Clapeyron equation for coexistence during the course of the simulations. The Clapeyron equation for equilibrium between two binary phases (e.g., liquid and solid) at constant pressure is given by
\begin{equation}
\label{eq:GibbsDuhem}
{ \frac {d \beta} {d \xi_B}  =  \frac { ( x_B^l -  x_B^s ) } {  \xi_B ( 1- \xi_B) (H^l - H^s ) } } , 
\end{equation}
where $\beta$ is the reciprocal temperature, $1/k_{B}T$, and $T$ the absolute temperature, $\xi_B$ the fugacity fraction of species $B$, $\xi_B = f_{B}/f_{A}$, $f_{i}$ the fugacity of species $i$ in solution, $x_B$ the mole fraction of species $B$, and $H$ is the molar enthalpy.  The right-hand side of Eq.~(\ref{eq:GibbsDuhem}) can be integrated numerically to find an equation for $\beta$ as a function of $\xi_B$ if we have an initial condition describing the temperature, fugacity fraction, enthalpies, and compositions at one coexistence point. 

The Gibbs-Duhem integration method is a more efficient approach to determining phase equilibrium in solid systems than the Gibbs-ensemble method,\cite{Frenkel:UnderstandingMS, Panagiotopoulos:GibbsEnsemble} because the important Monte Carlo move is changing the elemental identity of a particle rather than inserting or removing a particle from the system (inserting a particle is a low acceptance probability event). The Gibbs-Duhem integration  approach has been used to tackle a number of multicomponent, multiphase equilibrium problems including calculating the phase diagram of a binary Lennard-Jones fluid,\cite{Hitchcock:JCP} and calculating phase diagrams for colloids in polymer solutions.\cite{Kranendonk:HardSphere}  

In this paper, we determine the initial coexistence point by performing a microcanonical ensemble molecular dynamics simulation of an elemental system containing both a solid and liquid.\cite{Morris:CoexsitMD}  Using this approach, we directly measure the melting temperature of pure $A$ or $B$ (the solid and liquid fractions in the simulation cell evolve until the system reaches the equilibrium melting point).  Using this data as the starting point, we employ Kofke's Gibbs-Duhem integration technique\cite{Frenkel:UnderstandingMS, Hitchcock:JCP} within the framework of semigrand canonical Monte Carlo simulations.

\section{\label{sec:level4} Generic Behavior of  Binary Phase Diagrams for LJ-EAM materials}

In this section, we investigate solid/liquid phase diagrams for binary LJ-EAM systems. All of the temperature-composition phase diagrams reported below are for binary alloys at atmospheric pressure.  In particular, we focus on (a) cases where the melting points of the two elemental systems are identical and (b) cases in which the elemental systems have very different melting points. The former (similar melting points) is a relatively common case while the latter is more rare (although common for systems exhibiting liquid metal embrittlement.\cite{Joseph:LME})  In each case, we examine how variations of the ratio of Lennard-Jones diameters ${\sigma_{B}} / {\sigma_{A}}$ and well-depth ${\epsilon_{B}} / {\epsilon _{A}}$ affect the phase diagrams. We also compare major trends observed in the simulated phase diagrams with those measured experimentally (in order to evaluate the appropriateness of the LJ-EAM alloy model to mimic behavior in real metallic systems).

\subsection{\label{sec:sublevel4_1} Equal melting points}

\begin{figure}[!tbp]
\includegraphics[width=0.47\textwidth]{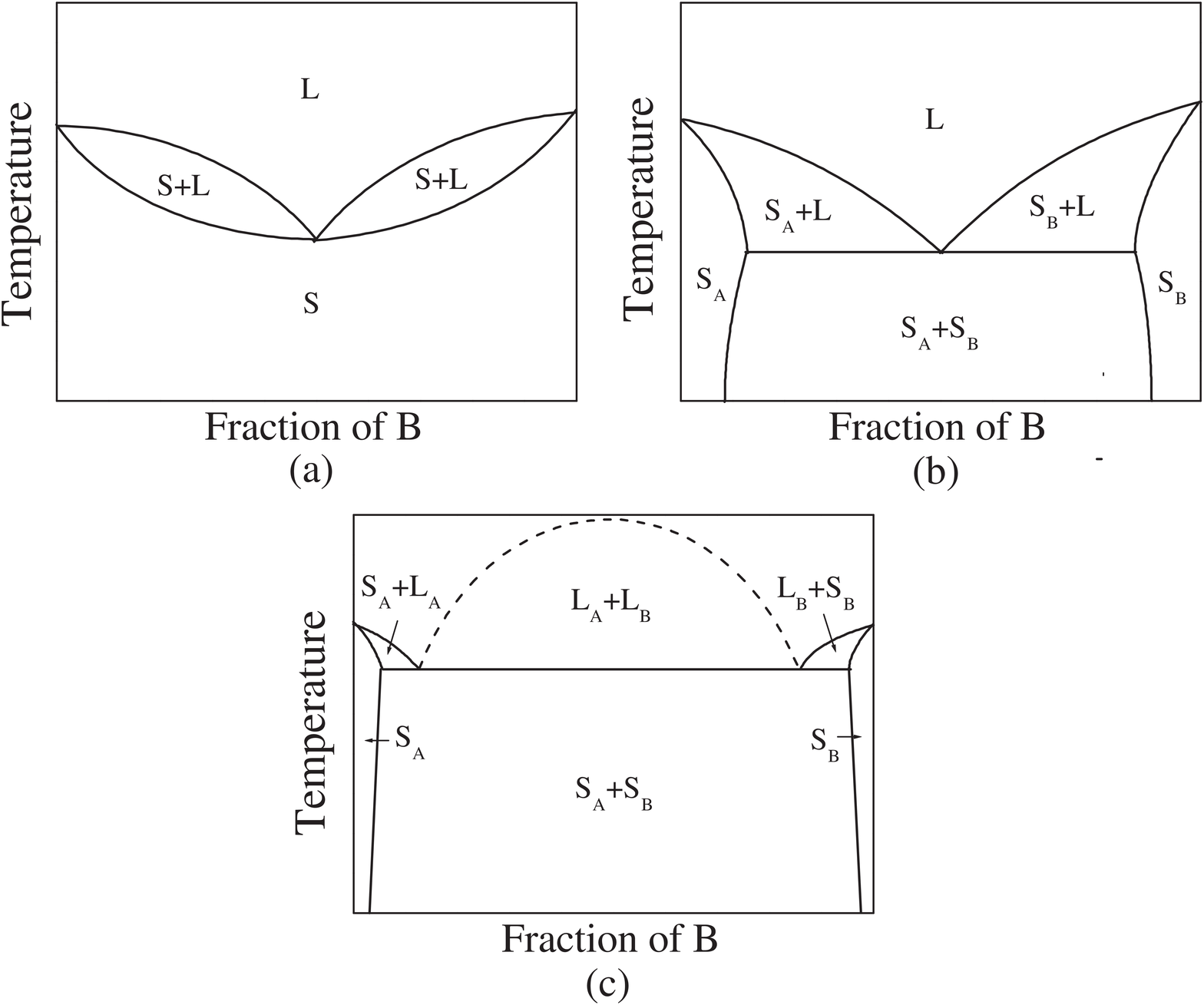} % Here is how to import EPS art
\caption
{ \label {fig:Schematic1}
Schematic phase diagrams for metallic binary alloys of comparable melting points: (a) azeotrope, (b) simple eutectic, and (c) a combination of a eutectic, a monotectic and a liquid phase miscibility gap. The symbols are as follows: $L$ refers to a liquid solution of $A$ and $B$, $S$ to a solid solution of $A$ and $B$, $S_A$ to a solid solution rich in $A$, and $S_B$ to a solid solution rich in $B$.  
}
\end{figure}

Figure~\ref{fig:Schematic1} shows three types of binary phase diagrams commonly observed in metallic alloys; we refer to these as azeotropes [Fig.~\ref{fig:Schematic1}(a)], simple eutectics [Fig.~\ref{fig:Schematic1}(b)], and liquid phase miscibility gap systems [Fig.~\ref{fig:Schematic1}(c)] of which there are several types.  We explicitly omit consideration of systems exhibiting compounds (for now). It is the atomic size mismatch and the cross-species pair interaction that determine which phase diagram type pertains.  To investigate the effect of both atomic size mismatch and cross-species pair interactions, we considered three series of cases: (1) like bonding ($\epsilon_{AB} = \epsilon_{A} = \epsilon_{B}$) with different atomic size ratios $\sigma_{B} / \sigma_{A}$, (2) equal atomic sizes ($ \sigma_{B} = \sigma_{A}=\sigma_{AB}$) and equal well-depths for the elements ($ \epsilon_{B} = \epsilon_{A}$) but vary the well-depth describing the $AB$ interactions $\epsilon_{AB}$, and (3) vary both $\sigma_{B} / \sigma_{A}$ and $\epsilon_{AB} / \epsilon_{A}$. The other parameters, $\widehat{A}$ and $\hat{\beta}$, are fixed at reference values 0.7 and 7 as mentioned in Sec.~\ref{sec:level2}

\begin{figure}[!tbp]
\includegraphics[width=0.40\textwidth]{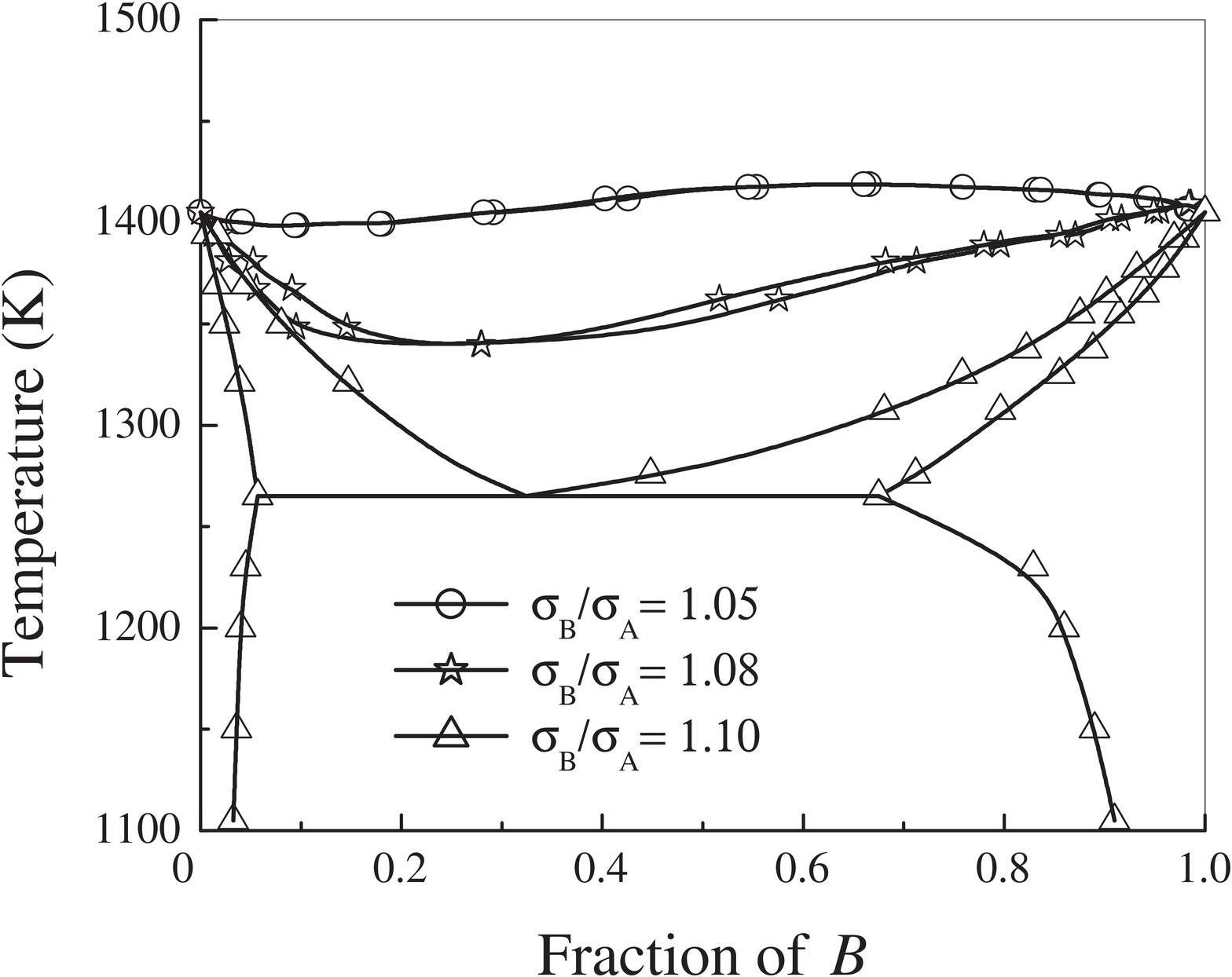} % Here is how to import EPS art
\caption
{ \label {fig:BMGReffect}
Solid-liquid phase diagrams with various atomic size differences.  Model alloys have the same melting point with zero heat of mixing ($\epsilon_{B} / \epsilon_{A} = 1.0$, $\epsilon_{AB} / \epsilon_{A} = 1.0$).
Solid-liquid phase diagrams were calculated with variation of atomic size difference ($\sigma_{B} / \sigma_{A} =1.05$, 1.08, and 1.1).  In this and subsequent phase diagrams, we estimate that each data point in the phase diagram has error bars of $\approx 10\,^{\circ}$ in temperature and $\approx 3 \%$ (although these errors have some variation depending on the relative stability of the different phases at each point along the phase boundaries).
}
\end{figure}

Figure~\ref{fig:BMGReffect} shows temperature-composition phase diagrams for a series of binary systems of type (1) (variation of atomic size).  When the atomic size difference is small (less than $8 \%$), the solid region of the phase diagram is a solution (for the temperature range examined).  However, as the atomic size difference increases, the degree of phase separation increases and the solid forms a miscibility gap that leads to a eutectic phase diagram starting at an atomic size difference between $8 \%$ and $10 \%$.  This cross-over from solid solution to eutectic phase diagrams occurs at a size difference that is much small than reported  for LJ\cite{Hitchcock:JCP} ($14 \sim 15 \%$) or hard sphere systems\cite{Kranendonk:HardSphere} ($12.5\%$). Moreover, the range of size difference for this type of simple eutectic phase diagram is very narrow;  further increase in this  difference leads to a miscibility gap in the liquid.  When a miscibility gap exists in the liquid, the phase diagram becomes more complex (similar to Fig.~\ref{fig:Schematic1}(c)), with negligible solubility in solid phases. Therefore, eutectic phase diagrams with deep eutectic point cannot be obtained by increasing size difference in LJ-EAM model.  This is not the case for LJ or hard sphere materials.

\begin{figure}[!tbp]
\includegraphics[width=0.40\textwidth]{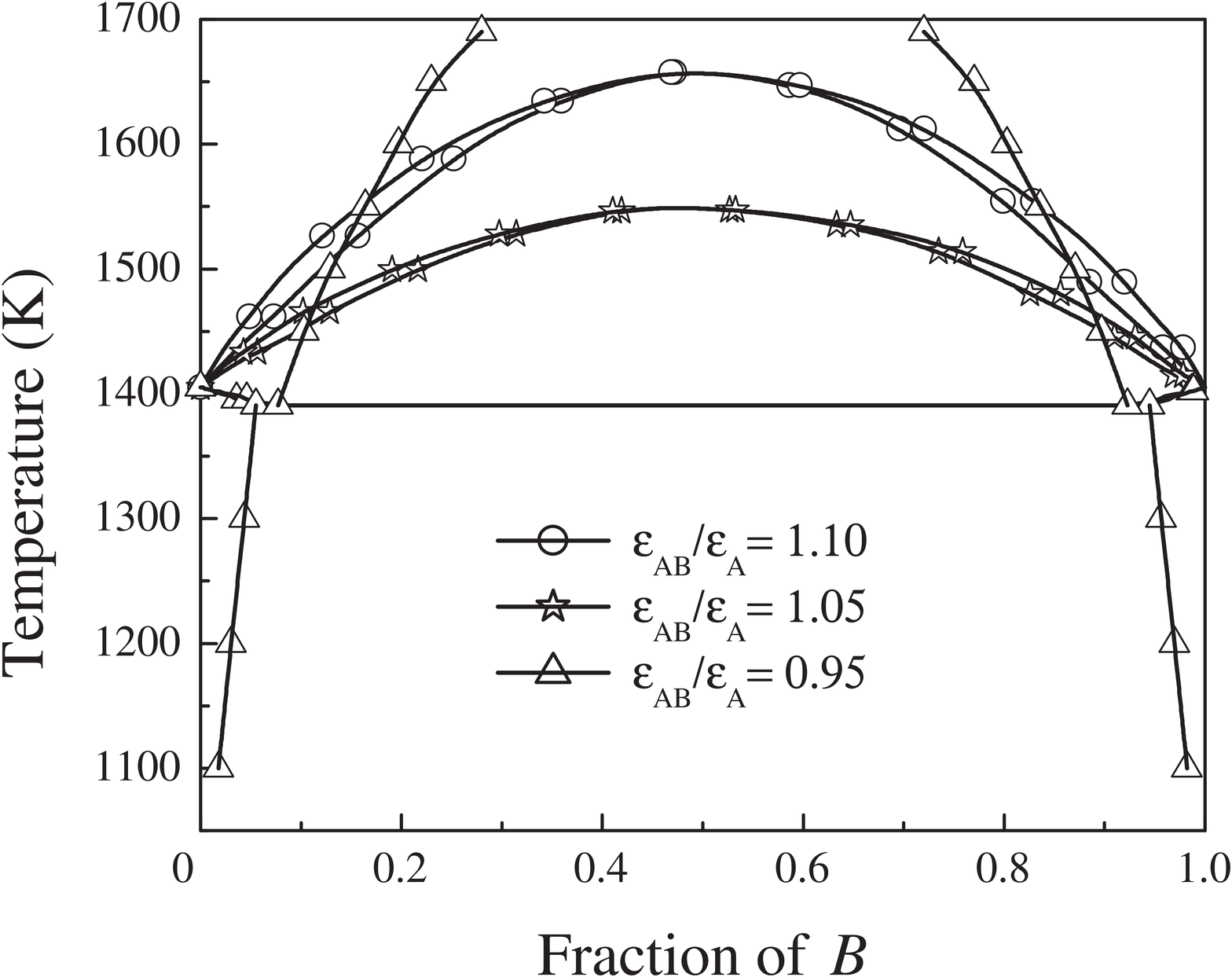} % Here is how to import EPS art
\caption
{ \label {fig:BMGDeffect}
Solid-liquid phase diagrams for different values of $\epsilon_{AB}$.  The pure metals, $A$ and $B$, have the same melting point and the same atomic size ($\epsilon_{B} / \epsilon_{A} = 1.0$, $\sigma_{B} / \sigma_{A} =1.0$).
The solid-liquid phase diagrams were calculated with different values of the cross-species interaction parameter ($\epsilon_{AB} / \epsilon_{A} =1.1$, 1.05, and 0.95).
}
\end{figure}

Figure~\ref{fig:BMGDeffect} shows temperature-composition phase diagrams for binary mixtures with same atomic size ($\sigma_{B}/\sigma_{A}=1$) and different values of cross-species interaction parameter. 
When $\epsilon_{AB} / \epsilon_{A} > 1$, heat of mixing is negative (exothermic solid solution) and A and B atoms ``like each other''. In these cases, the phase diagrams form continuous solid solutions over the whole composition range and the liquidus appears to be parabolic.  The maximum temperature for which the solid and liquid coexist increases as $\epsilon_{AB} / \epsilon_{A}$ increases (the heat of mixing is more negative the larger the $\epsilon_{AB} / \epsilon_{A}$ ratio).
For a binary mixture with well-depth ratio of unity $\epsilon_{AB} / \epsilon_{A} = 1$, all of the atoms are indistinguishable - hence, the phase diagram would simply be a horizontal line at $T=1405$ K.
On the other hand, when $\epsilon_{AB} / \epsilon_{A} < 1$, the heat of mixing is positive (endothermic solid solution) and A-B atom ``dislike each other''. Under this circumstance, the liquidus curve is  concave for  $1 < \epsilon_{AB} / \epsilon_{A} < 0.95$ (not shown). 
For $\epsilon_{AB} / \epsilon_{A} \le 0.95$, the two species are no longer miscible at all compositions and a large miscibility gap appears in both the solid and liquid phase regions, as shown in the phase diagram in Fig.~\ref{fig:BMGDeffect}.

\begin{figure}[!tbp]
\includegraphics[width=0.40\textwidth]{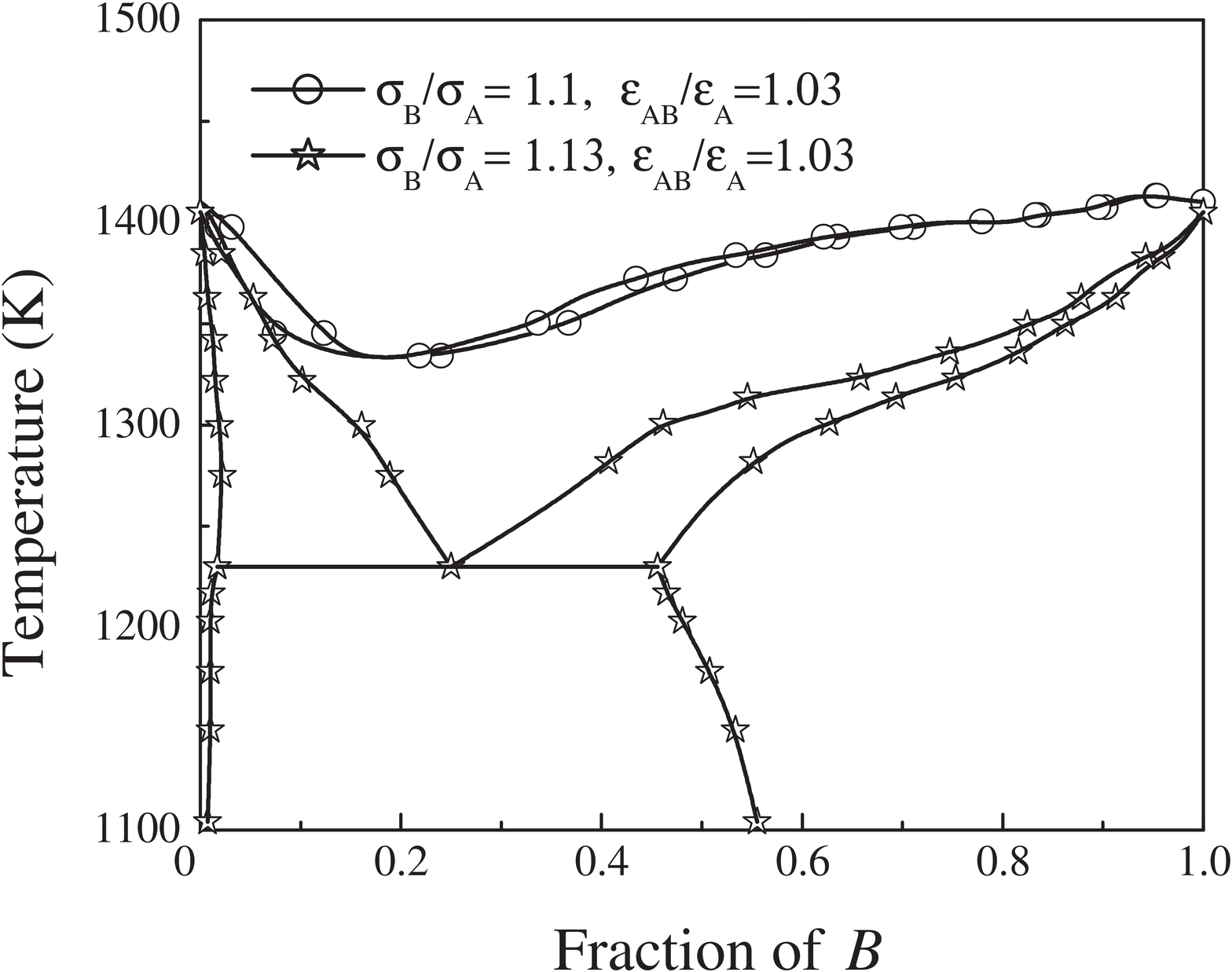} % Here is how to import EPS art
\caption
{ \label {fig:BMGoffset}
Solid-liquid phase diagram for a model binary alloys with atomic size differences of $10\%$ - $13\%$  and $\epsilon_{AB} / \sigma_{A}=1.03$.
}
\end{figure}

These results suggest that the heat of mixing alone does not control the melting point (liquidus).  Rather, atomic size difference also plays an important role, especially for forming eutectic phase diagrams.  However, both the size difference and  heat of mixing affect  the tendency towards mixing and can act to compensate each other. Therefore, controlling atomic size difference together with  the heat of mixing can lead to a wide range of types of binary phase diagrams.  For example, increasing the liquidus temperature by increasing $\epsilon_{AB} / \epsilon_{A}$ can be compensated by increasing the atomic size difference.  When large atomic size difference are combined with relatively large $\epsilon_{AB} / \epsilon_{A}$,  eutectic phase diagrams with deep eutectics can be formed, as shown in Fig.~\ref{fig:BMGoffset}.
This trend is quite interesting because it seems to be related with the rule of thumb for making metallic glasses.  In metallic glass systems, it has been established, empirically, that the ability to form glasses is greatest in  multicomponent systems in which the atomic size difference is large and the heat of mixing is strongly negative (large $\epsilon_{AB} / \sigma_{A}$).\cite{Inoue:BMGs}  The present phase diagram results suggest that this is also the description of the condition for the formation of deep eutectics.  Deep eutectics are also known as systems for which glass formation is particularly easy.

\subsection{\label{sec:sublevel4_2} Phase diagrams of large melting point difference}

\begin{figure}[!tbp]
\includegraphics[width=0.48\textwidth]{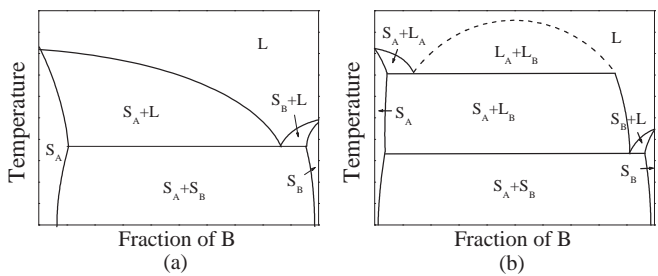} % Here is how to import EPS art
\caption
{ \label {fig:Schematic2}
Schematic phase diagrams of binary alloys where the two components have very different melting points.  Two such cases are  (a) a eutectic with significant solubility in the solid and (b) a diagram with two eutectic points and a miscibility gap in the liquid phase.
}
\end{figure}

Solid-liquid coexistence is a key to many materials processing strategies involving solidification and in-service conditions where a solid metal is in contact with another, liquid metal.  In many of the latter cases, the liquid phases consist of low melting point species, such as Hg, Ga, Bi, Pb, and Sn.  Metallic binary systems in which the melting points of the two components differ greatly typically show one of two types of simple solid/liquid phase diagram (provided no intermetallic compounds form): these are eutectics phase diagrams with or without a liquid phase miscibility gap, as shown in Fig.~\ref{fig:Schematic2}. When there is a liquid phase miscibility gap, the solubility of the minority species in the solid phase is usually very small [although it often appears exaggerated in schematic phase diagrams, such as Fig.~\ref{fig:Schematic2}(b)].  However, in other binary systems, such as Al-Ga and Zn-Ga, the  solubility in the solid phase can be significant over the entire temperature range, in spite of large melting point difference [see Fig.~\ref{fig:Schematic2}(a)]. 

\begin{figure}[!tbp]
\includegraphics[width=0.40\textwidth]{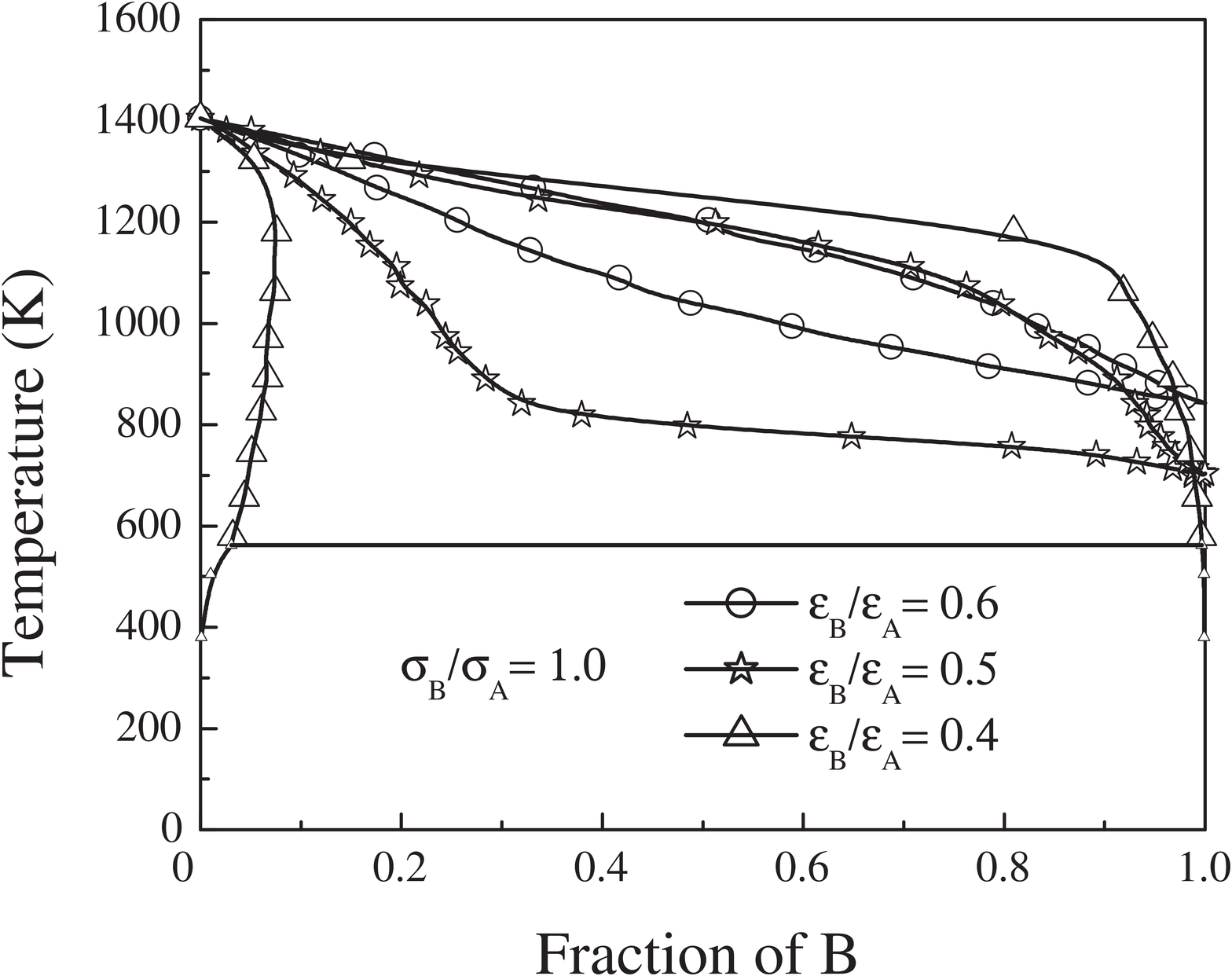} % Here is how to import EPS art
\caption
{ \label {fig:LMEEeffect}
Solid-liquid phase diagrams for binary alloys with different melting points of component $B$, as controlled by different choices of $\epsilon_{B} / \epsilon_{A} (=$0.6, 0.5, and 0.4).  These alloys have the same atomic sizes ($\sigma_{B} / \sigma_{A} = 1.0$, and the parameter $\epsilon_{AB}$ is described by Eq.~(\ref{eq:NewRule})).
}
\end{figure}

Phase diagrams were calculated for several different binary systems, where we varied the melting point of species $B$.  Figure~\ref{fig:LMEEeffect} shows the temperature-composition phase diagrams for systems with the atom size ratio fixed as $\sigma_{B} / \sigma_{A} = $1 and well-depth ratios of $\epsilon_{B} / \epsilon_{A} = 0.6$, 0.5, and 0.4. When the melting point of $B$ is comparable to $A$, the system forms a  solid solution with a spindle shaped solid-liquid two-phase region.  As $\epsilon_{B} / \epsilon_{A}$ decreases, $A$-$B$ become weaker and the phase diagram evolves to a  eutectic diagram. It is difficult to see the solid-liquid two-phase region at the $B$-rich side of the phase diagram in Fig.~\ref{fig:LMEEeffect} because the eutectic point is close to pure $B$.  Nonetheless, we assure the reader that this is indeed a eutectic, just like in Fig.~\ref{fig:Schematic2}(a).  As $ \epsilon_{B} / \epsilon_{A} $ decreases further, the solubility of $B$  in the solid phase and the solubility of $A$ in the liquid phase decrease. This trend agrees with the observation that eutectic phase diagrams determined from experiment tend to show smaller solubilities as the ratio of the melting points of the two species deviates further from unity. 

\begin{figure}[!tbp]
\includegraphics[width=0.40\textwidth]{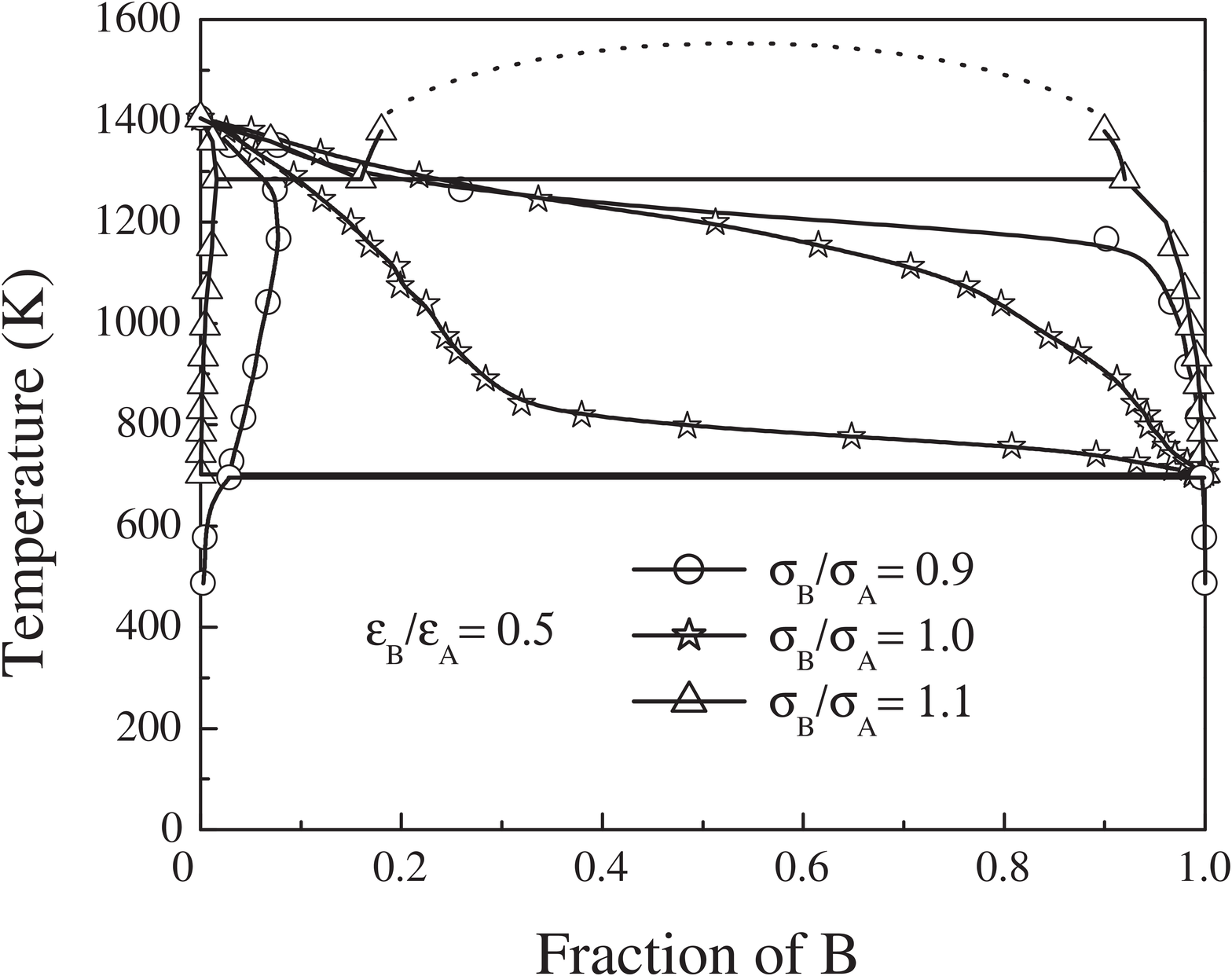} % Here is how to import EPS art
\caption
{ \label {fig:LMEReffect}
Solid-liquid phase diagrams with different atomic sizes ratios ($\sigma_{B} / \sigma_{A} =$0.9, 1.0, and 1.1). The  melting point of $B$ was held constant by fixing $\epsilon_{B} / \epsilon_{A} = 0.5$ and $\widehat{A}_{B}=0.7$).
}
\end{figure}

Phase diagrams were also determined for binary systems with different atomic size ratios.  Figure~\ref{fig:LMEReffect} shows the temperature-composition phase diagrams for  binary mixtures with a fixed well-depth ratio, $ \epsilon_{B} / \epsilon_{A} = 0.5$, and several diameter ratios, $ \sigma_{B} / \sigma_{A} = 0.9$, 1.0, and 1.1. When there is no atomic size mismatch, the solubilities in the solid  and liquid phases are quite large, despite the large melting point difference.  But, as the atomic size mismatch decreases, these solubilities decrease. Interestingly, when the atomic size of $B$ is larger than that of $A$, the size effect is dominant (i.e., the solubility is negligible over the entire temperature range and a miscibility gap appears in liquid phase). 
\begin{figure}[!tbp]
\includegraphics[width=0.40\textwidth]{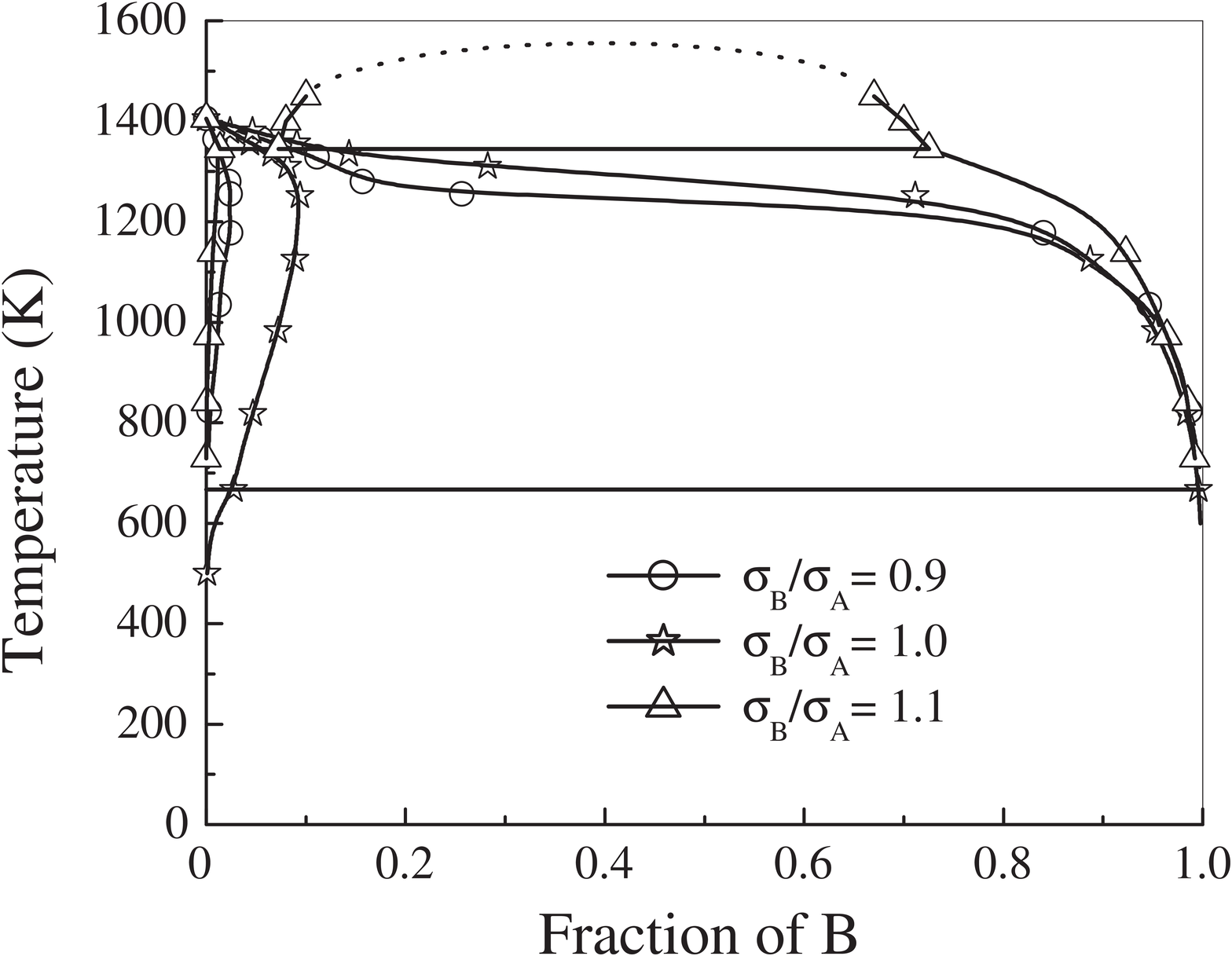} % Here is how to import EPS art
\caption
{ \label {fig:LMEAeffect}
Solid-liquid phase diagrams with different atomic sizes ratios ($\sigma_{B} / \sigma_{A} =$0.9, 1.0, and 1.1). The  melting point of $B$ was held constant by fixing $\widehat{A}_{B}=0.9$ and $\epsilon_{B} / \epsilon_{A} = 1$.
}
\end{figure}

The melting point of $B$ can be set by the choice of the many-body interaction parameter $\widehat{A}_{B}$ (in addition to choosing the well-depth ratio $ \epsilon_{B} / \epsilon_{A}$). The melting point of $B$ decreases with increasing  $\widehat{A}_{B}$ even though the cohesive energy remains unchanged. Figure~\ref{fig:LMEAeffect} shows the solid-liquid phase diagram with $\widehat{A}_{B}=0.9$ for several different atomic size ratios. Effect of atomic size mismatch is still valid for this kind of solid/liquid pairs. The trends in the phase diagrams with atomic size mismatch are similar in this case to those shown in Fig.~\ref{fig:LMEReffect}.

\subsection{\label{sec:sublevel4_3} Comparison with real binary alloy systems}

Since the phase diagrams shown above were determined within the framework of generic interatomic potentials LJ-EAM, it is interesting to inquire to what degree choosing parameters in the potential can lead to phase diagrams that are consistent with those found experimentally in real metallic systems. We can compare the simulation data to experimental results to verify the ability of the LJ-EAM model to mimic behavior in real metallic systems.
The elements found in column IB of the periodic table, copper, silver, and gold, are common FCC metals that are well described with the embedded atom method framework.  We determine the LJ-EAM parameters, $\epsilon$, $\sigma$ and $\widehat{A}$, to reproduce the cohesive energy, lattice parameter and melting temperatures of these column IB elements (see Ref.~\onlinecite{Baskes:LJEAMPRSL}), as shown in Table~\ref{tab:table2}.
\begin{table}[!tbp]
\caption
{\label{tab:table2}
Cohesive energy $E_{0}$, lattice parameter $a_{0}$, and melting point $T_{\rm{m}}$ for column IB FCC metals\cite{Kittel:SolidStatePhys, Smith:MetalRef} and the corresponding potential parameters $\epsilon$, $\sigma$, and $\widehat{A}$.
}
\begin{ruledtabular}
\begin{tabular}{lcccccc}
element & $E_{0}$ (eV) & $a_{0}$ (\AA) & $T_{\rm{m}}$(K) & $\epsilon$ & $\sigma$ & $\widehat{A}$ \\
\hline
Cu & 3.54 & 3.62 & 1357 & 0.59  & 2.277 & 0.7  \\
Ag & 2.85 & 4.09 & 1235 & 0.475 & 2.573 & 0.66 \\
Au & 3.93 & 4.08 & 1392 & 0.655 & 2.570 & 0.8  \\
\end{tabular}
\end{ruledtabular}
\end{table}
\begin{figure}[!tbp]
\includegraphics[width=0.35\textwidth]{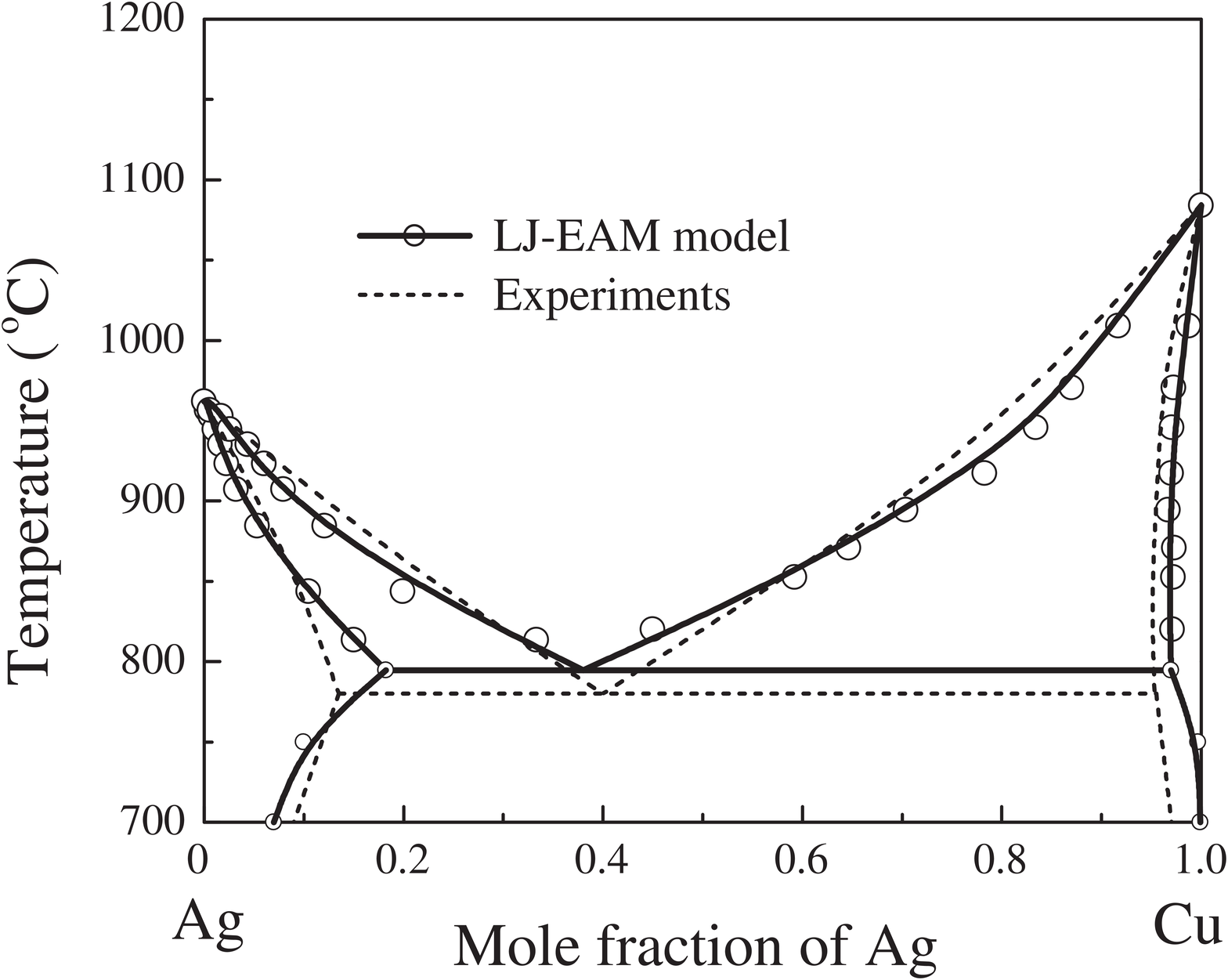} % Here is how to import EPS art
\includegraphics[width=0.35\textwidth]{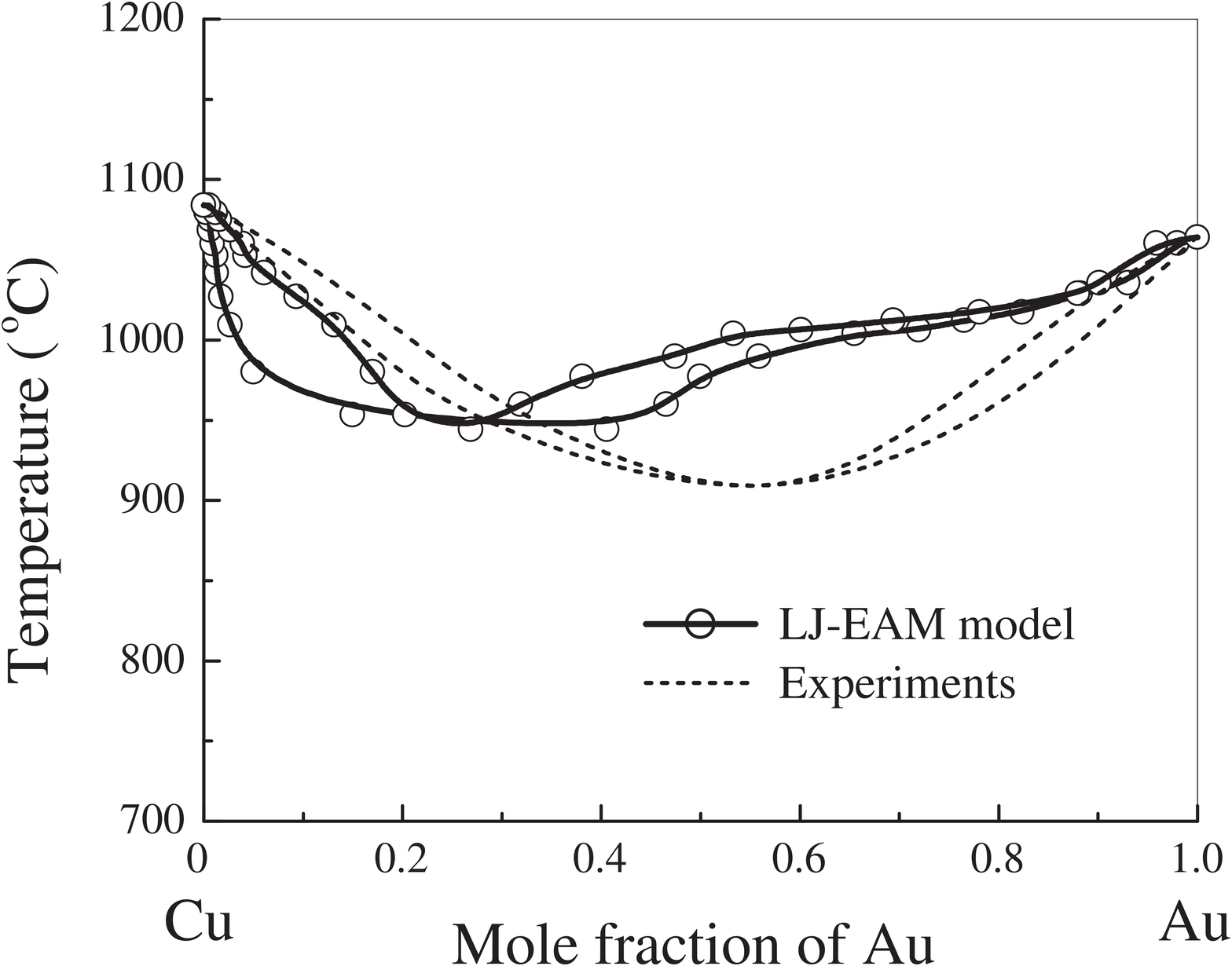} % Here is how to import EPS art
\includegraphics[width=0.35\textwidth]{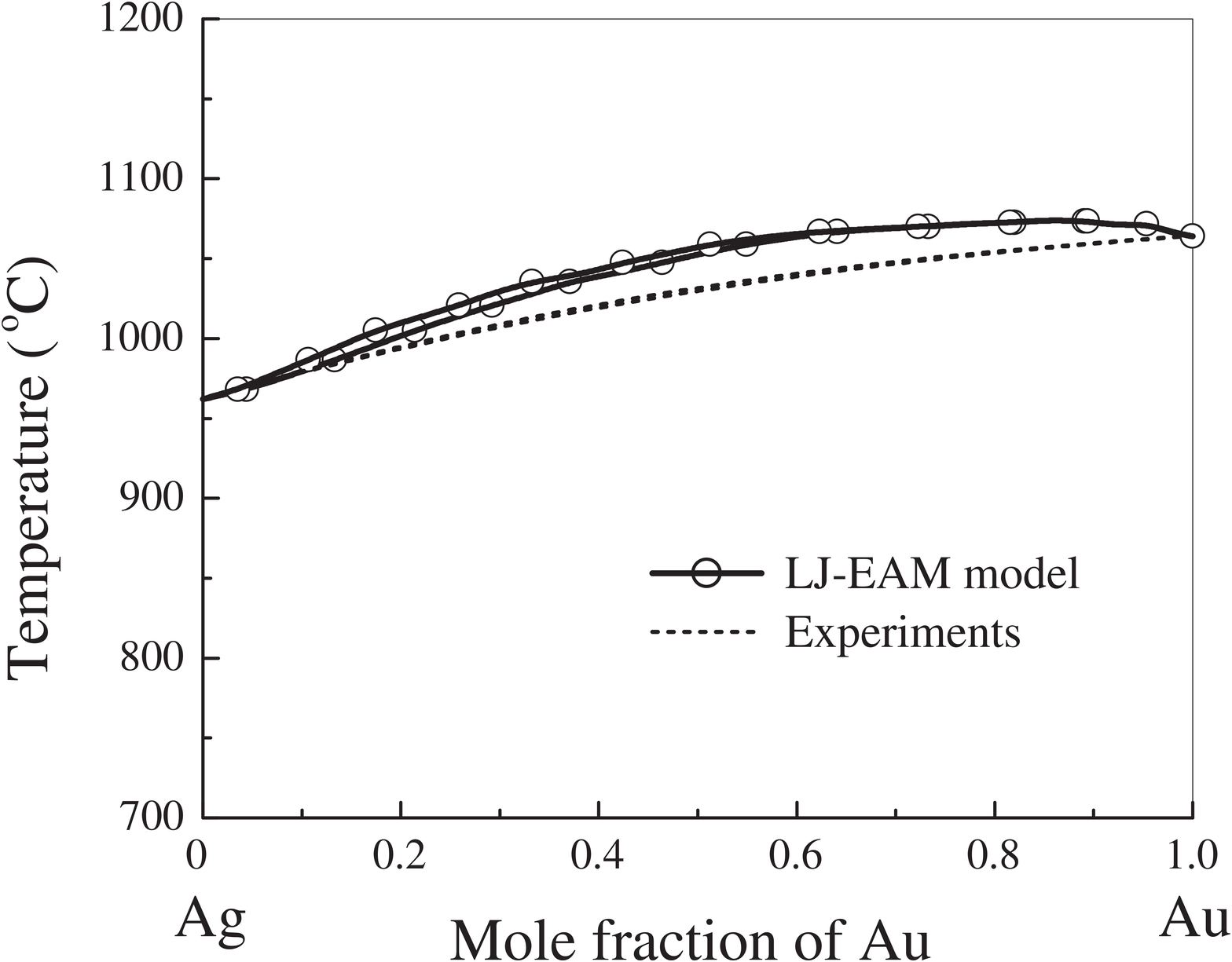} % Here is how to import EPS art
\caption
{ \label {fig:AgCuAu}
Solid-liquid phase diagrams for (a) Ag-Cu, (b) Cu-Au, and (c) Ag-Au. Cohesive energy and LJ diameter were fitted to real properties of materials and $\epsilon_{AB} / \epsilon^0_{AB}$ was adjusted in LJ-EAM binary alloy model. $\epsilon_{AB} / \epsilon^0_{AB}=1.03$ for Ag-Cu, $\epsilon_{AB} / \epsilon^0_{AB}=1.03$ for Cu-Au, $\epsilon_{AB} / \epsilon^0_{AB}=1.0$ for Ag-Au, where $\epsilon^0_{AB}$ is given in Eq.~(\ref{eq:NewRule}).
}
\end{figure}

We employ these parameters within the  LJ-EAM framework to calculate the corresponding Ag-Cu, Cu-Au, and Ag-Au binary phase diagrams. By only adjusting the ratio $\epsilon_{AB} / \epsilon_{A}$ [note, the final values were very close to those predicted by Eq.~(\ref{eq:NewRule})], we are able to reasonably reproduce the experimental phase diagrams (each with a unique topology). Although the agreement certainly is not perfect, the phase diagram type,  the temperature ranges of the features of the diagrams, and solubilities are in good agreement with experiment.  This is remarkable given that only one parameter was varied (and the atomic sizes and cohesive energies are available from experiment). This type of agreement is possible for many binary systems, provided that they do not exhibit intermetallic compounds.

\section{\label{sec:level5} Conclusion}

We developed an approach to determining LJ-EAM potentials for alloys and used these to determine the solid-liquid phase diagrams for binary metallic alloys using Kofke's Gibbs-Duhem integration technique combined with semigrand canonical Monte Carlo simulations.  Inclusion of many-body interactions led to phase diagrams which can be quite different from those determined using LJ or hard sphere materials. We demonstrated that it is possible to produce a wide-range of experimentally observed binary phase diagrams (with no intermetallic phases) by reference to the atomic sizes and cohesive energies of the two elemental materials and by judicious choice of a single parameter that controls the pairwise interactions of these two elements.  In addition, we provided a formula that leads to good choices for this one free parameter.  Within this framework, we performed a systematic investigation of the effect of relative atomic sizes and cohesive energies of the elements on the binary phase diagrams.  Finally, we demonstrated that this approach leads to good agreement with several experimental binary phase diagrams.  The main benefit of this approach is not, in our opinion, to accurately reproduce the phase diagrams of real materials.  Rather, it is to provide a method by which material properties can be continuously changed in simulation studies in order to develop understanding of mechanisms and properties in a manner not available to experiment. To this end, the relationship between the adjustable potentials and the phase diagrams they imply is central.

%%%%%%%%%%%%%%%%%%%%%%%%%%%%%%% Body of Paper %%%%%%%%%%%%%%%%%%%%%%%%%%%%%%%%%%

%-------------------- Acknowledgments ---------------------%

\begin{acknowledgments}
The authors gratefully acknowledge the support of Korea Science \& Engineering Foundation (H.-S. N) and the support of the US Department of Energy, Office of Fusion Energy Sciences, Grant No. DE-FG02-011ER54628.
\end{acknowledgments}

%\newpage %Just because of unusual number of tables stacked at end

%-------------------- Acknowledgments ---------------------%

\appendix

\section{Derivation of $\phi_{AB} (r)$}

To derive formulation for $\phi(r)$, $L1_0$ structure with c/a=1 was considered as reference state.  In LJ-EAM formalism, the energy per atom of this structure as a function of dilation is given by 
\begin{eqnarray}
\label{eq:A1}
E^{\textrm {LJ-EAM}}_{L1_0} (r) =&& \frac {1} {2} \left [ F_{A} (\bar \rho_{A}^{L1_0} (r) ) + F_{B} (\bar \rho_{B}^{L1_0} (r) ) \right ]
\nonumber\\
&&+ \left [ \phi_{AA} (r) + 4 \phi_{AB} (r) + \phi_{BB} (r) \right ]
\nonumber\\
&&+ \frac {3} {2} \left [ \phi_{AA} (ar) + \phi_{BB} (ar) \right ], 
\end{eqnarray}
where
\begin{eqnarray}
\bar \rho_{A}^{L1_0} (r) = \frac {1} {3} \left [ 2 \rho_{B} (r) + \rho_{A} (r) \right ] + \frac {1} {2} \rho_{A} (ar)
\end{eqnarray}
and
\begin{eqnarray}
\bar \rho_{B}^{L1_0} (r) = \frac {1} {3} \left [ 2 \rho_{A} (r) + \rho_{B} (r) \right ] + \frac {1} {2} \rho_{B} (ar).
\end{eqnarray}
If we rewrite the energy of this system in terms of LJ pair potentials,
\begin{eqnarray}
\label{eq:A2}
E^{\textrm {LJ}}_{L1_0} (r) =&& \phi_{AA}^{LJ} (r) + 4 \phi_{AB}^{LJ} (r) + \phi_{BB}^{LJ} (r). 
\end{eqnarray}
By setting $E^{\textrm {LJ-EAM}}_{L1_0} (r)= E^{\textrm {LJ}}_{L1_0} (r)$, we can derive Eq.~(\ref{eq:EAMphiHetro}).

\section{Mixing Rule for $\epsilon_{AB}$}
In this appendix, we show the derivation of the expression we employ for describing the interaction between unlike species.  Given Eq.~(\ref{eq:EAMphiHetro}), this reduces to the determination of $\phi^{LJ}_{AB}$.  The length scale parameter $\sigma^{LJ}_{AB}$ is simply the arithmetic average of those for the elements.  Therefore, we only require a mixing rule for the well depth parameter $\epsilon_{AB}$.  In this appendix, we show the origin of the choice for this parameter that was quoted in Eq.~(\ref{eq:NewRule}).  

In order to keep this analysis simple, we rewrite Eq.~(\ref{eq:A1}) for the special case of nearest-neighbor interactions, $\sigma_{A}=\sigma_{B}$ and $\hat{\beta}_{A}=\hat{\beta}_{B}$:
\begin{eqnarray}
\label{eq:A3}
E_{L1_0} (r) =&& \frac {1} {2} \left [ F_{A} (\bar \rho_{A}^{0} (r) ) + F_{B} (\bar \rho_{B}^{0} (r) ) \right ]
\nonumber\\
&&+ \left [ \phi_{AA} (r) + 4 \phi_{AB} (r) + \phi_{BB} (r) \right ]
\nonumber\\
= && \phi_{AA}^{LJ} (r) + 4 \phi_{AB}^{LJ} (r) + \phi_{BB}^{LJ} (r). 
\end{eqnarray}

In equilibrium at zero pressure, $r=r_{eq}(=\sqrt[6]{2} \sigma_{A}=\sqrt[6]{2} \sigma_{B})$, $\rho_{A}^{0} (r_{eq})=\rho_{B}^{0} (r_{eq})=1$, $\phi_{AA}^{LJ} (r_{eq})=\epsilon_{A}$, $\phi_{BB}^{LJ} (r_{eq})=\epsilon_{B}$, and $\phi_{AB}^{LJ} (r_{eq})=\epsilon_{AB}$.  Substituting these expressions into Eqs.~(\ref{eq:EAMf}) and (\ref{eq:EAMphiHomo}), yields
\begin{eqnarray}
&&F_{A} (\bar \rho_{A}^{0} (r_{eq}) ) = ~ \frac {1} {2} \widehat{A}_{A} Z_{1} \epsilon_{A} = 6 \widehat{A}_{A} \epsilon_{A},
\nonumber\\
&&F_{B} (\bar \rho_{B}^{0} (r_{eq}) ) = ~ \frac {1} {2} \widehat{A}_{B} Z_{1} \epsilon_{B} = 6 \widehat{A}_{B} \epsilon_{B},
\end{eqnarray}
and
\begin{eqnarray}
\phi_{AA} (r_{eq}) &&= \phi_{AA}^{LJ} (r_{eq}) - \left ( \frac {2} {Z_1} \right ) F_{A} (\bar \rho_{A}^{0} (r_{eq}) )
\nonumber\\
&&=\epsilon_{A} - \widehat{A}_{A} \epsilon_{A} = \left (1-\widehat{A}_{A} \right ) \epsilon_{A},
\nonumber\\
\phi_{BB} (r_{eq}) &&= \left (1-\widehat{A}_{B} \right ) \epsilon_{B}. 
\end{eqnarray}
The energy per atom at $r$=$r_{eq}$ then becomes
\begin{eqnarray}
E_{L1_0} (r_{eq}) =&& \frac {1} {2} \left [ 6 \widehat{A}_{A} \epsilon_{A} + 6 \widehat{A}_{B} \epsilon_{B} \right ]
\nonumber\\
&&+ \left [ (1-\widehat{A}_{A})\epsilon_{A} + 4 \phi_{AB} (r) + (1-\widehat{A}_{B})\epsilon_{B} \right ]
\nonumber\\
= && \epsilon_{A} + 4 \epsilon_{AB} + \epsilon_{B}. 
\end{eqnarray}
Solving this for $\epsilon_{AB}$ yields 
\begin{eqnarray}
\label{eq:AeAB}
\epsilon_{AB} = 
\frac {1} {2} \left ( \widehat{A}_{A} \epsilon_{A} + \widehat{A}_{B} \epsilon_{B} \right ) 
+ \phi_{AB} (r_{eq})
\end{eqnarray}
Now, determining $\epsilon_{AB}$ is a matter of determining $\phi_{AB} (r_{eq})$.
As described in the text, we obtain $\phi_{AB} (r_{eq})$ by applying the Lorentz-Bertholet mixing rule to $\phi (r_{eq})$ rather than to $\epsilon$ (these are equivalent for the pairwise potentials):
\begin{eqnarray}
\phi_{AB} (r_{eq}) = && \sqrt { |\phi_{AA} (r_{eq}) \phi_{BB} (r_{eq})| }
\nonumber\\
= && \sqrt { |(1 - \widehat{A}_{A}) \epsilon_{A} (1 - \widehat{A}_{B}) \epsilon_{B}| }.
\end{eqnarray}
Substituting this into Eq.~(\ref{eq:AeAB}) gives the modified mixing rule, Eq.~(\ref{eq:NewRule}):
\begin{eqnarray}
\epsilon_{AB} = 
&&\frac {1} {2} \left ( \widehat{A}_{A} \epsilon_{A} + \widehat{A}_{B} \epsilon_{B} \right ) 
\nonumber\\
&&+ \sqrt { |(1 - \widehat{A}_{A}) \epsilon_{A} (1 - \widehat{A}_{B}) \epsilon_{B}| }.
\end{eqnarray}

%%%%%%%%%%%%%%%%%%%%%%%%%%%%%%% BIBLIOGRAPHY %%%%%%%%%%%%%%%%%%%%%%%%%%%%%%%%%%%

% \bibliography{Ih_PRL}% Produces the bibliography via BibTeX.

\bibliography{PhaseDiagram}        % qhe.bib is the name of our database

\begin{thebibliography}{32}
\expandafter\ifx\csname natexlab\endcsname\relax\def\natexlab#1{#1}\fi
\expandafter\ifx\csname bibnamefont\endcsname\relax
  \def\bibnamefont#1{#1}\fi
\expandafter\ifx\csname bibfnamefont\endcsname\relax
  \def\bibfnamefont#1{#1}\fi
\expandafter\ifx\csname citenamefont\endcsname\relax
  \def\citenamefont#1{#1}\fi
\expandafter\ifx\csname url\endcsname\relax
  \def\url#1{\texttt{#1}}\fi
\expandafter\ifx\csname urlprefix\endcsname\relax\def\urlprefix{URL }\fi
\providecommand{\bibinfo}[2]{#2}
\providecommand{\eprint}[2][]{\url{#2}}

\bibitem[{\citenamefont{Ohno et~al.}(1999)\citenamefont{Ohno, Esfarjani, and
  Kawazoe}}]{Ohno:CMS}
\bibinfo{author}{\bibfnamefont{K.}~\bibnamefont{Ohno}},
  \bibinfo{author}{\bibfnamefont{K.}~\bibnamefont{Esfarjani}},
  \bibnamefont{and} \bibinfo{author}{\bibfnamefont{Y.}~\bibnamefont{Kawazoe}},
  \emph{\bibinfo{title}{Computational Materials Science}}
  (\bibinfo{publisher}{Springer}, \bibinfo{address}{Berlin},
  \bibinfo{year}{1999}).

\bibitem[{\citenamefont{Frenkel and Smit}(2002)}]{Frenkel:UnderstandingMS}
\bibinfo{author}{\bibfnamefont{D.}~\bibnamefont{Frenkel}} \bibnamefont{and}
  \bibinfo{author}{\bibfnamefont{B.}~\bibnamefont{Smit}},
  \emph{\bibinfo{title}{Understanding Molecular Simulation}}
  (\bibinfo{publisher}{Academic Press}, \bibinfo{address}{San Diego},
  \bibinfo{year}{2002}), \bibinfo{edition}{2nd} ed.

\bibitem[{\citenamefont{Lennard-Jones}(1924)}]{LennardJones:LJpotential}
\bibinfo{author}{\bibfnamefont{J.~E.} \bibnamefont{Lennard-Jones}},
  \bibinfo{journal}{Proc. R. Soc. Lond. A} \textbf{\bibinfo{volume}{106}},
  \bibinfo{pages}{463} (\bibinfo{year}{1924}).

\bibitem[{\citenamefont{Morse}(1929)}]{Morse:Morsepotential}
\bibinfo{author}{\bibfnamefont{P.~M.} \bibnamefont{Morse}},
  \bibinfo{journal}{Phys. Rev.} \textbf{\bibinfo{volume}{34}},
  \bibinfo{pages}{57} (\bibinfo{year}{1929}).

\bibitem[{\citenamefont{Jensen et~al.}(1973)\citenamefont{Jensen, Kristensen,
  and Cotterill}}]{Jensen:LJexample}
\bibinfo{author}{\bibfnamefont{E.~J.} \bibnamefont{Jensen}},
  \bibinfo{author}{\bibfnamefont{W.~D.} \bibnamefont{Kristensen}},
  \bibnamefont{and} \bibinfo{author}{\bibfnamefont{R.~M.~J.}
  \bibnamefont{Cotterill}}, \bibinfo{journal}{Philos. Mag.}
  \textbf{\bibinfo{volume}{27}}, \bibinfo{pages}{623} (\bibinfo{year}{1973}).

\bibitem[{\citenamefont{Broughton and Gilmer}(1983)}]{Broughton:LJexample}
\bibinfo{author}{\bibfnamefont{J.~Q.} \bibnamefont{Broughton}}
  \bibnamefont{and} \bibinfo{author}{\bibfnamefont{G.~H.}
  \bibnamefont{Gilmer}}, \bibinfo{journal}{J. Chem. Phys}
  \textbf{\bibinfo{volume}{79}}, \bibinfo{pages}{5095} (\bibinfo{year}{1983}).

\bibitem[{\citenamefont{Vitek and Srolovitz}(1989)}]{Vitek:BeyondPair}
\bibinfo{editor}{\bibfnamefont{V.}~\bibnamefont{Vitek}} \bibnamefont{and}
  \bibinfo{editor}{\bibfnamefont{D.~J.} \bibnamefont{Srolovitz}}, eds.,
  \emph{\bibinfo{title}{Atomistic Simulation of Materials Beyond Pair
  Potential}} (\bibinfo{publisher}{Plenum}, \bibinfo{address}{New York},
  \bibinfo{year}{1989}).

\bibitem[{\citenamefont{Carlsson}(1990)}]{Carlsson:BeyondPair}
\bibinfo{author}{\bibfnamefont{A.~E.} \bibnamefont{Carlsson}},
  \emph{\bibinfo{title}{Beyond Pair Potentials in Elemental Transition Metals
  and Semiconductors}} (\bibinfo{publisher}{Academic Press},
  \bibinfo{address}{Boston}, \bibinfo{year}{1990}).

\bibitem[{\citenamefont{Daw and Baskes}(1983)}]{Daw:EAM}
\bibinfo{author}{\bibfnamefont{M.~S.} \bibnamefont{Daw}} \bibnamefont{and}
  \bibinfo{author}{\bibfnamefont{M.~I.} \bibnamefont{Baskes}},
  \bibinfo{journal}{Phys. Rev. Lett.} \textbf{\bibinfo{volume}{50}},
  \bibinfo{pages}{1285} (\bibinfo{year}{1983}).

\bibitem[{\citenamefont{Holian et~al.}(1991)\citenamefont{Holian, Voter,
  Wagner, Ravelo, Chen, Hoover, Hoover, Hammerberg, and Dontje}}]{Holian:LJEAM}
\bibinfo{author}{\bibfnamefont{B.~L.} \bibnamefont{Holian}},
  \bibinfo{author}{\bibfnamefont{A.~F.} \bibnamefont{Voter}},
  \bibinfo{author}{\bibfnamefont{N.~J.} \bibnamefont{Wagner}},
  \bibinfo{author}{\bibfnamefont{R.~J.} \bibnamefont{Ravelo}},
  \bibinfo{author}{\bibfnamefont{S.~P.} \bibnamefont{Chen}},
  \bibinfo{author}{\bibfnamefont{W.~G.} \bibnamefont{Hoover}},
  \bibinfo{author}{\bibfnamefont{C.~G.} \bibnamefont{Hoover}},
  \bibinfo{author}{\bibfnamefont{J.~E.} \bibnamefont{Hammerberg}},
  \bibnamefont{and} \bibinfo{author}{\bibfnamefont{T.~D.}
  \bibnamefont{Dontje}}, \bibinfo{journal}{Phys. Rev. A}
  \textbf{\bibinfo{volume}{43}}, \bibinfo{pages}{2655} (\bibinfo{year}{1991}).

\bibitem[{\citenamefont{Baskes}(1999)}]{Baskes:LJEAMPRL}
\bibinfo{author}{\bibfnamefont{M.~I.} \bibnamefont{Baskes}},
  \bibinfo{journal}{Phys. Rev. Lett.} \textbf{\bibinfo{volume}{83}},
  \bibinfo{pages}{2592} (\bibinfo{year}{1999}).

\bibitem[{\citenamefont{Srinivasan and Baskes}(2004)}]{Baskes:LJEAMPRSL}
\bibinfo{author}{\bibfnamefont{S.~G.} \bibnamefont{Srinivasan}}
  \bibnamefont{and} \bibinfo{author}{\bibfnamefont{M.~I.}
  \bibnamefont{Baskes}}, \bibinfo{journal}{Proc. R. Soc. Lond. A}
  \textbf{\bibinfo{volume}{460}}, \bibinfo{pages}{1649} (\bibinfo{year}{2004}).

\bibitem[{\citenamefont{Baskes and Stan}(2003)}]{Baskes:PhaDiaEAM}
\bibinfo{author}{\bibfnamefont{M.~I.} \bibnamefont{Baskes}} \bibnamefont{and}
  \bibinfo{author}{\bibfnamefont{M.}~\bibnamefont{Stan}},
  \bibinfo{journal}{Metall. Mater. Trans. A} \textbf{\bibinfo{volume}{34}},
  \bibinfo{pages}{435} (\bibinfo{year}{2003}).

\bibitem[{\citenamefont{Joseph et~al.}(1999)\citenamefont{Joseph, Picat, and
  Barbier}}]{Joseph:LME}
\bibinfo{author}{\bibfnamefont{B.}~\bibnamefont{Joseph}},
  \bibinfo{author}{\bibfnamefont{M.}~\bibnamefont{Picat}}, \bibnamefont{and}
  \bibinfo{author}{\bibfnamefont{F.}~\bibnamefont{Barbier}},
  \bibinfo{journal}{Eur. Phys. J. Appl. Phys.} \textbf{\bibinfo{volume}{5}},
  \bibinfo{pages}{19} (\bibinfo{year}{1999}).

\bibitem[{\citenamefont{Kranendonk and Frenkel}(1991)}]{Kranendonk:HardSphere}
\bibinfo{author}{\bibfnamefont{W.~G.~T.} \bibnamefont{Kranendonk}}
  \bibnamefont{and} \bibinfo{author}{\bibfnamefont{D.}~\bibnamefont{Frenkel}},
  \bibinfo{journal}{Mol. Phys.} \textbf{\bibinfo{volume}{72}},
  \bibinfo{pages}{679} (\bibinfo{year}{1991}).

\bibitem[{\citenamefont{Vlot et~al.}(1997)\citenamefont{Vlot, van Miltenburg,
  Oonk, and van~der Eerden}}]{Vlot:PhaDiaLJ}
\bibinfo{author}{\bibfnamefont{M.~J.} \bibnamefont{Vlot}},
  \bibinfo{author}{\bibfnamefont{J.~C.} \bibnamefont{van Miltenburg}},
  \bibinfo{author}{\bibfnamefont{H.~A.~J.} \bibnamefont{Oonk}},
  \bibnamefont{and} \bibinfo{author}{\bibfnamefont{J.~P.} \bibnamefont{van~der
  Eerden}}, \bibinfo{journal}{J. Chem. Phys} \textbf{\bibinfo{volume}{107}},
  \bibinfo{pages}{10102} (\bibinfo{year}{1997}).

\bibitem[{\citenamefont{Hitchcock and Hall}(1999)}]{Hitchcock:JCP}
\bibinfo{author}{\bibfnamefont{M.~R.} \bibnamefont{Hitchcock}}
  \bibnamefont{and} \bibinfo{author}{\bibfnamefont{C.~K.} \bibnamefont{Hall}},
  \bibinfo{journal}{J. Chem. Phys} \textbf{\bibinfo{volume}{110}},
  \bibinfo{pages}{11433} (\bibinfo{year}{1999}).

\bibitem[{\citenamefont{Octoby}(1991)}]{Octoby:DFT}
\bibinfo{author}{\bibfnamefont{W.~G.~T.} \bibnamefont{Octoby}},
  \bibinfo{journal}{Mol. Phys.} \textbf{\bibinfo{volume}{72}},
  \bibinfo{pages}{679} (\bibinfo{year}{1991}).

\bibitem[{\citenamefont{Kofke}(1993{\natexlab{a}})}]{Kofke:GibbsDuhemMolPhys}
\bibinfo{author}{\bibfnamefont{D.~A.} \bibnamefont{Kofke}},
  \bibinfo{journal}{Mol. Phys.} \textbf{\bibinfo{volume}{78}},
  \bibinfo{pages}{1331} (\bibinfo{year}{1993}{\natexlab{a}}).

\bibitem[{\citenamefont{Kofke}(1993{\natexlab{b}})}]{Kofke:GibbsDuhemJCP}
\bibinfo{author}{\bibfnamefont{D.~A.} \bibnamefont{Kofke}},
  \bibinfo{journal}{J. Chem. Phys} \textbf{\bibinfo{volume}{98}},
  \bibinfo{pages}{4149} (\bibinfo{year}{1993}{\natexlab{b}}).

\bibitem[{\citenamefont{Rowlinson}(1982)}]{Rowlinson:LBrule}
\bibinfo{author}{\bibfnamefont{J.~S.} \bibnamefont{Rowlinson}},
  \emph{\bibinfo{title}{Liquids and Liquid Mixtures}}
  (\bibinfo{publisher}{Butterworth Scientific}, \bibinfo{address}{London},
  \bibinfo{year}{1982}).

\bibitem[{\citenamefont{Yurtsever and F.Calvo}(2000)}]{Yurtsever:LJEAMcluster}
\bibinfo{author}{\bibfnamefont{E.}~\bibnamefont{Yurtsever}} \bibnamefont{and}
  \bibinfo{author}{\bibnamefont{F.Calvo}}, \bibinfo{journal}{Phys. Rev. B}
  \textbf{\bibinfo{volume}{62}}, \bibinfo{pages}{9977} (\bibinfo{year}{2000}).

\bibitem[{\citenamefont{Wang et~al.}(2002)\citenamefont{Wang, Zhou, and
  Liu}}]{Wang:LJEAMmelting}
\bibinfo{author}{\bibfnamefont{T.}~\bibnamefont{Wang}},
  \bibinfo{author}{\bibfnamefont{F.~X.} \bibnamefont{Zhou}}, \bibnamefont{and}
  \bibinfo{author}{\bibfnamefont{Y.~W.} \bibnamefont{Liu}},
  \bibinfo{journal}{Chinese Physics} \textbf{\bibinfo{volume}{11}},
  \bibinfo{pages}{139} (\bibinfo{year}{2002}).

\bibitem[{\citenamefont{Baskes}(2004)}]{Baskes:LJEAMbinary}
\bibinfo{author}{\bibfnamefont{M.~I.} \bibnamefont{Baskes}},
  \bibinfo{journal}{JOM} \textbf{\bibinfo{volume}{56}}, \bibinfo{pages}{45}
  (\bibinfo{year}{2004}).

\bibitem[{\citenamefont{Pao and Srolovitz}(2006)}]{Srolovitz:StressMorphology}
\bibinfo{author}{\bibfnamefont{C.-W.} \bibnamefont{Pao}} \bibnamefont{and}
  \bibinfo{author}{\bibfnamefont{D.~J.} \bibnamefont{Srolovitz}},
  \bibinfo{journal}{Phys. Rev. Lett.} \textbf{\bibinfo{volume}{96}},
  \bibinfo{pages}{186103} (\bibinfo{year}{2006}).

\bibitem[{\citenamefont{Pryde}(1969)}]{Pryde:LiquidState}
\bibinfo{author}{\bibfnamefont{J.~A.} \bibnamefont{Pryde}},
  \emph{\bibinfo{title}{The Liquid State}} (\bibinfo{publisher}{Hutchinson
  University Library}, \bibinfo{address}{London}, \bibinfo{year}{1969}).

\bibitem[{\citenamefont{Rose et~al.}(1984)\citenamefont{Rose, Smith, Guinea,
  and Ferrante}}]{Rose:EAMUniversalFeatures}
\bibinfo{author}{\bibfnamefont{J.~H.} \bibnamefont{Rose}},
  \bibinfo{author}{\bibfnamefont{J.~R.} \bibnamefont{Smith}},
  \bibinfo{author}{\bibfnamefont{F.}~\bibnamefont{Guinea}}, \bibnamefont{and}
  \bibinfo{author}{\bibfnamefont{J.}~\bibnamefont{Ferrante}},
  \bibinfo{journal}{Phys. Rev. B} \textbf{\bibinfo{volume}{29}},
  \bibinfo{pages}{2963} (\bibinfo{year}{1984}).

\bibitem[{\citenamefont{Panggiotopoulos
  et~al.}(1988)\citenamefont{Panggiotopoulos, Quirke, Stapleton, and
  Tildesley}}]{Panagiotopoulos:GibbsEnsemble}
\bibinfo{author}{\bibfnamefont{A.~Z.} \bibnamefont{Panggiotopoulos}},
  \bibinfo{author}{\bibfnamefont{N.}~\bibnamefont{Quirke}},
  \bibinfo{author}{\bibfnamefont{M.}~\bibnamefont{Stapleton}},
  \bibnamefont{and} \bibinfo{author}{\bibfnamefont{D.~J.}
  \bibnamefont{Tildesley}}, \bibinfo{journal}{Mol. Phys.}
  \textbf{\bibinfo{volume}{63}}, \bibinfo{pages}{527} (\bibinfo{year}{1988}).

\bibitem[{\citenamefont{Morris et~al.}(1994)\citenamefont{Morris, Wang, Ho, and
  Chan}}]{Morris:CoexsitMD}
\bibinfo{author}{\bibfnamefont{J.~R.} \bibnamefont{Morris}},
  \bibinfo{author}{\bibfnamefont{C.~Z.} \bibnamefont{Wang}},
  \bibinfo{author}{\bibfnamefont{K.~M.} \bibnamefont{Ho}}, \bibnamefont{and}
  \bibinfo{author}{\bibfnamefont{C.~T.} \bibnamefont{Chan}},
  \bibinfo{journal}{Phys. Rev. B} \textbf{\bibinfo{volume}{49}},
  \bibinfo{pages}{3109} (\bibinfo{year}{1994}).

\bibitem[{\citenamefont{Inoue}(2000)}]{Inoue:BMGs}
\bibinfo{author}{\bibfnamefont{A.}~\bibnamefont{Inoue}}, \bibinfo{journal}{Acta
  Mater.} \textbf{\bibinfo{volume}{48}}, \bibinfo{pages}{279}
  (\bibinfo{year}{2000}).

\bibitem[{\citenamefont{Kittel}(1995)}]{Kittel:SolidStatePhys}
\bibinfo{author}{\bibfnamefont{C.}~\bibnamefont{Kittel}},
  \emph{\bibinfo{title}{Introduction to Solid State Physics}}
  (\bibinfo{publisher}{Wiley}, \bibinfo{address}{New York},
  \bibinfo{year}{1995}), \bibinfo{edition}{7th} ed.

\bibitem[{\citenamefont{Smith}(1976)}]{Smith:MetalRef}
\bibinfo{editor}{\bibfnamefont{C.~J.} \bibnamefont{Smith}}, ed.,
  \emph{\bibinfo{title}{Metal Reference Book}}
  (\bibinfo{publisher}{Butterworths}, \bibinfo{address}{London},
  \bibinfo{year}{1976}), \bibinfo{edition}{5th} ed.

\end{thebibliography}

\end{document}